\documentclass[journal]{IEEEtran}

\usepackage{cite}
\usepackage{float}
\usepackage{bibcheck}
\usepackage{graphicx}
\usepackage[justification=centering]{caption}
\usepackage{amssymb}
\usepackage{amsopn}
\usepackage{amstext}
\usepackage{lscape}
\usepackage[justification=centering]{subcaption}
\usepackage{multirow}
\usepackage{color}
\usepackage[backref]{hyperref}
\newcommand{\reffig}[1]{Figure \ref{#1}}
\newcommand{\reftab}[1]{Table \ref{#1}}
\newcommand{\refsec}[1]{Section \ref{#1}}
\newcommand{\refequ}[1]{Equation \ref{#1}}
\newcommand{\tabincell}[2]{\begin{tabular}{@{}#1@{}}#2\end{tabular}}

\begin{document}
	\title{Deep Feature Mining via Attention-based BiLSTM-GCN for Human Motor Imagery Recognition}
	\author{Yimin~Hou, Shuyue~Jia,~\IEEEmembership{Student Member,~IEEE}, Xiangmin~Lun, Shu~Zhang, Tao~Chen, Fang~Wang, Jinglei~Lv
	\IEEEcompsocitemizethanks{
		\IEEEcompsocthanksitem This work was supported by the National Natural Science Foundation of China under Grant 31772059.
		\IEEEcompsocthanksitem Yimin Hou, Xiangmin Lun, Shuyue Jia, Tao Chen, and Fang Wang are with the School of Automation Engineering, Northeast Electric Power University, Jilin, China.
		\IEEEcompsocthanksitem Xiangmin Lun is also with the College of Mechanical and Electric Engineering, Changchun University of Science and Technology, Changchun, China.
		\IEEEcompsocthanksitem Shu Zhang is with the School of Computer Science, Northwestern Poly-technical University, Xi’an, China.
		\IEEEcompsocthanksitem Jinglei Lv is with School of Biomedical Engineering \& Brain and Mind Center, University of Sydney, Sydney, NSW, Australia.
	}
}

\markboth{Journal of \LaTeX\ Class Files,~Vol.~14, No.~8, December~2021}
{Shell \MakeLowercase{\textit{et al.}}: Bare Demo of IEEEtran.cls for Computer Society Journals}

\IEEEtitleabstractindextext{
\begin{abstract}
Recognition accuracy and response time are both critically essential ahead of building practical electroencephalography (EEG) based brain-computer interface (BCI). However, recent approaches have either compromised in the classification accuracy or responding time. This paper presents a novel deep learning approach designed towards both remarkably accurate and responsive motor imagery (MI) recognition based on scalp EEG. Bidirectional Long Short-term Memory (BiLSTM) with the Attention mechanism is employednvolutional neural network (GCN) promotes the decoding performance by cooperating with the topological structure of features, which are estimated from the overall data. Particularly, this method is trained and tested on the short EEG recording with only 0.4 second length, and the result has shown effective and efficient prediction based on individual and group-wise training, with 98.81\% and 94.64\% accuracy, respectively, which outperformed all the state-of-the-art studies. The introduced deep feature mining approach can precisely recognize human motion intents from raw and almost instant EEG signals, which paves the road to translate the EEG based MI recognition to practical BCI systems.
\end{abstract}

}

\maketitle
\IEEEdisplaynontitleabstractindextext

\section{Introduction}\label{Introduction}
\IEEEPARstart{R}ecently, brain-computer interface (BCI) plays a promising role in assisting and rehabilitating patients from paralysis, epilepsy, and brain injuries via interpreting neural activities to control the peripherals\cite{bouton2016restoring, schwemmer2018meeting}. Among the non-invasive brain activity acquisition systems, electroencephalography (EEG)-based BCI gains extensive attention recently given its higher temporal resolution and portability. Hence, it has been popularly employed to assist the recovery of patients from motor impairments, e.g., amyotrophic lateral sclerosis (ALS), spinal cord injury (SCI), or stroke survivors\cite{daly2008brain, pereira2018eeg}. Specifically, researchers have focused on the recognition of motor imagery (MI) based on EEG and translating the brain activities into specific motor intentions. In such a way, users can further manipulate external devices or exchange information with the surroundings\cite{pereira2018eeg}. Although researchers have developed several MI-based prototype applications, there is still space of improvement before the practical clinical translation could be promoted\cite{schwemmer2018meeting, mahmood2019fully}. De facto, to achieve effective and efficient control via only MI, both precise EEG decoding and quick response are eagerly expected. However, few existing works of literature are competent in both perspectives. In this study, we explore the possibility of a deep learning framework to tackle the challenge.

\begin{figure*}[h]
	\centering
	\includegraphics[width=\linewidth]{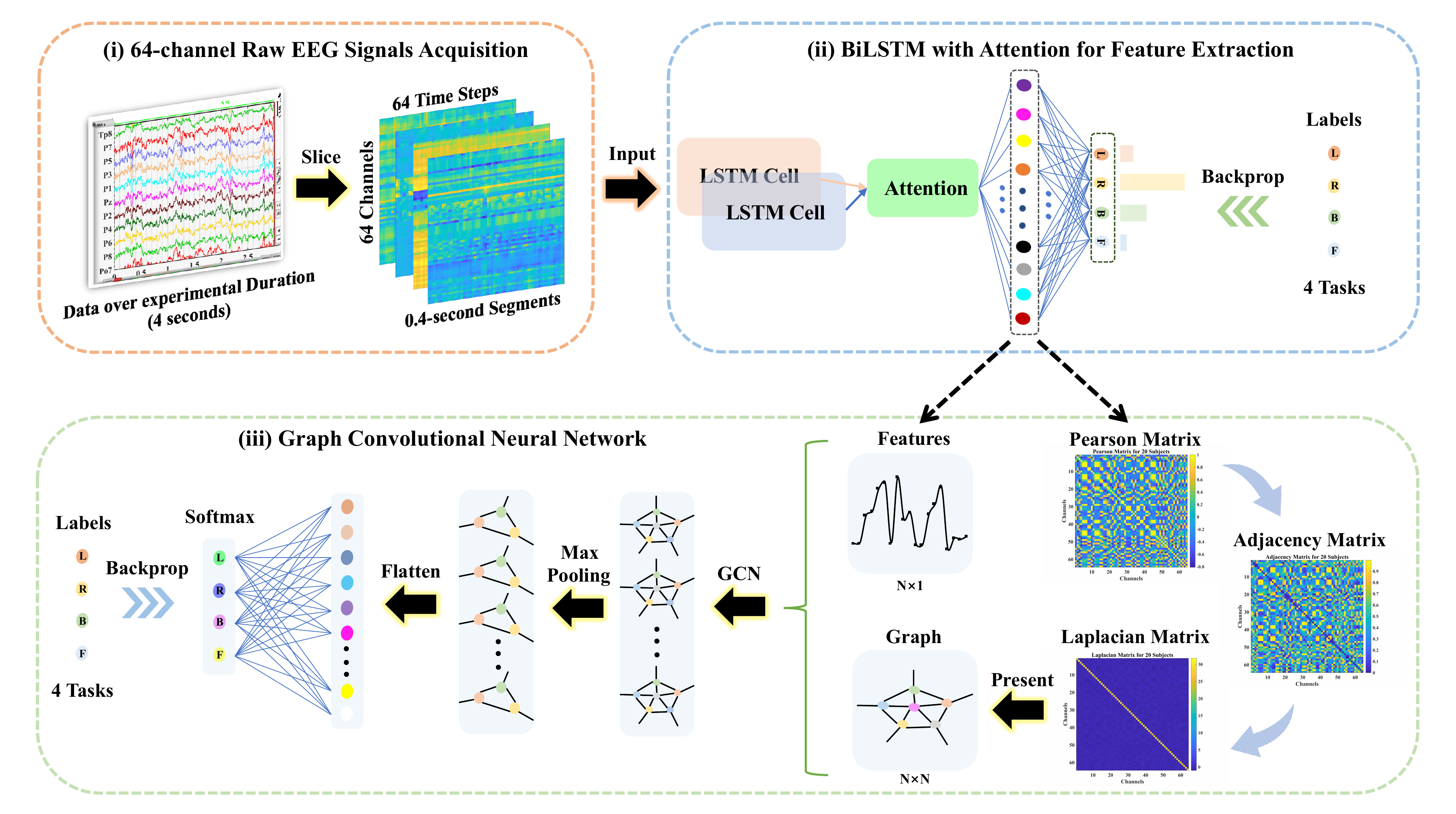}
	\caption{The schematical overview consisted of 64-channel raw EEG signals acquisition, the BiLSTM with Attention model for feature extraction, and the GCN model for classification.}
	\label{Framework}
\end{figure*}

	\subsection{Related Work}\label{Related Work}
	Lately, Deep Learning (DL) attracted increasing attention in many disciplines because of its promising performance in classification tasks\cite{lecun2015deep}. A growing number of works have shown that DL will play a pivotal role in the precise decoding of brain activities\cite{schwemmer2018meeting}. Especially, recent works have been carried out on EEG motion intention detection. A primary current focus is to implement the DL-based approach to decode EEG MI tasks, which have attained promising results\cite{lotte2018review}. Due to the high temporal resolution of EEG signals, methods related to the recurrent neural network (RNN)\cite{rumelhart1986learning}, which can analyze time-series data, were extensively applied to filter and classify EEG sequences, i.e., time points\cite{zhang2018spatial, wang2018lstm, luo2018exploring, guler2005recurrent, zhang2018mindid}. In reference\cite{zhang2018spatial}, a novel RNN framework with spatial and temporal filtering was put forward to classify EEG signals for emotion recognition, and achieved 95.4\% accuracy for three classes with a 9 seconds' segment as a sample. Wang \emph{et~al.} and Luo \emph{et~al.} performed Long Short-term Memory (LSTM)\cite{hochreiter1997long} to handle time slices' signals, and achieved 77.30\% and 82.75\% accuracy, respectively\cite{wang2018lstm, luo2018exploring}. Reference \cite{zhang2018mindid} presented Attention-based RNN for EEG-based person identification, which attained 99.89\% accuracy for 8 participants at the subject level with 4 seconds' signals as a sample. However, it can be found that in these studies, signals over experimental duration were recognized as samples, which resulted in a slow responsive prediction.
	
	Apart from RNN, Convolutional Neural Network (CNN)\cite{fukushima1980neocognitron, lecun1998gradient} has been performed to decode EEG signals as well\cite{hou2019novel, dose2018end}. Hou \emph{et~al.} proposed ESI and CNN, and achieved competitive results, i.e., 94.50\% and 96.00\% accuracy at the group and subject levels, respectively, for four-class classification. What is more, by combining CNN with graph theory, the Graph Convolutional Neural Network (GCN)\cite{henaff2015deep, bruna2014spectral, duvenaud2015convolutional, niepert2016learning, defferrard2016convolutional} approach was presented lately, taking consideration of the functional topological relationship of EEG electrodes\cite{wang2018eeg, zhang2019gcb, song2018eeg, wang2019phase}. In references\cite{wang2018eeg} and \cite{zhang2019gcb}, a GCN with broad learning approach was proposed, and attained 93.66\% and 94.24\% accuracy, separately, for EEG emotion recognition. Song \emph{et~al.} and Wang \emph{et~al.} introduced dynamical GCN (90.40\% accuracy) and phase-locking value-based GCN (84.35\% accuracy) to recognize different emotions\cite{song2018eeg, wang2019phase}. Highly accurate prediction has been accomplished via the GCN model. But few researchers have investigated the approach in the area of EEG MI decoding. 
	
	\begin{figure*}[h]
		\centering
		\includegraphics[width=\linewidth]{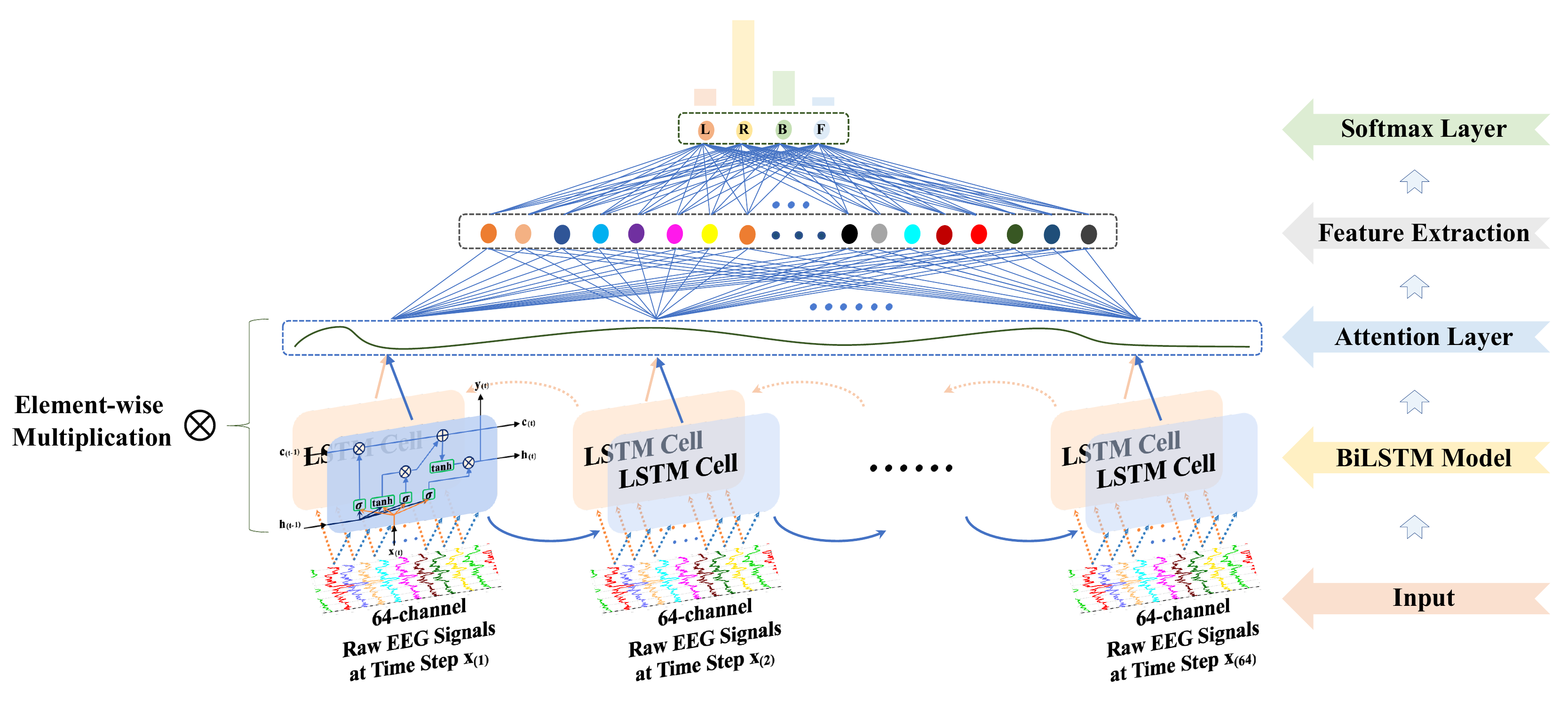}
		\caption{Presented BiLSTM with the Attention mechanism for feature extraction.}
		\label{BiLSTM-with-Attention}
	\end{figure*}

	\subsection{Contribution of This Paper}\label{Contribution of This Paper}
	Towards accurate and fast MI recognition, an Attention-based BiLSTM-GCN was introduced to mine the effective features from raw EEG signals. The main contributions were summarized as follows.\\
	\textbf{(i)} To our best knowledge, this work was the first that combined BiLSTM with the GCN to decode EEG tasks. \\
	\textbf{(ii)} The Attention-based BiLSTM successfully derived relevant features from raw EEG signals. Followed by the GCN model, it enhanced the decoding performance by considering the internal topological structure of features.\\
	\textbf{(iii)} The proposed feature mining approach managed to decode EEG MI signals with stably reproducible results yielding remarkable robustness and adaptability, even countering considerable inter-trial and inter-subject variability.
	
	\subsection{Organization of This Paper}\label{Organization of This Paper}
	The rest of this paper was organized in the following. The preliminary knowledge of the BiLSTM, Attention mechanism and GCN was introduced in \refsec{Methodology}, which was the foundation of the presented approach. In \refsec{Results and Discussion}, experimental details and numerical results were presented, followed by the conclusion in \refsec{Conclusion}. 
	
\section{Methodology}\label{Methodology}
	\subsection{Pipeline Overview}\label{Pipeline Overview}
	The framework of the proposed method was demonstrated in \reffig{Framework}. \\
	\textbf{(i)} 64-channel raw EEG signals were acquired via the BCI2000, and then the 4 seconds' (experimental duration) signals were sliced into 0.4-second segments over time, where the dimension of each segment was 64 channels $\times$ 64 time steps. \\
	\textbf{(ii)} The Attention-based BiLSTM was put forward to filter 64-channel (spatial information) and 0.4-second (temporal information) raw EEG data and derived features from the fully-connected neurons. \\
	\textbf{(iii)} The Pearson, Adjacency, and Laplacian Matrices of overall features were introduced sequentially to represent the topological structure of features, i.e., as a graph. Followed by the features and its corresponding graph representation as the input, the GCN model was performed to classify four-class MI tasks. 
	
	\subsection{BiLSTM with Attention}\label{BiLSTM with Attention}
		\subsubsection{BiLSTM Model}\label{BiLSTM Model}
		RNN-based approaches have been extensively applied to analyze EEG time-series signals. An RNN cell, though alike a feedforward neural network, has connections pointing backward, which sends its output back to itself. The learned features of an RNN cell at time step $t$ are influenced by not only input signals $\mathbf{x}_{(t)}$, but also the output (state) at time step $t$-1. This design mechanism dictates that RNN-based methods can handle sequential data, e.g., time points signals, by unrolling the network through time. The LSTM and Gated Recurrent Unit (GRU)\cite{choetal2014learning} are the most popular variants of the RNN-based approaches. In \refsec{Proposed Approach}, the paper compared the performance of the welcomed models experimentally, and the BiLSTM with Attention displayed in \reffig{BiLSTM-with-Attention} outperformed others due to better detection of the long-term dependencies of raw EEG signals. 
		\begin{equation}
			\label{LSTM-1}
			\mathbf{i}_{(t)}=\sigma\left(\mathbf{W}_{x i}^{T} \cdot \mathbf{x}_{(t)}+\mathbf{W}_{h i}^{T} \cdot \mathbf{h}_{(t-1)}+\mathbf{b}_{i}\right)
		\end{equation}
		\begin{equation}
			\mathbf{f}_{(t)}=\sigma\left(\mathbf{W}_{x f}^{T} \cdot \mathbf{x}_{(t)}+\mathbf{W}_{h f}^{T} \cdot \mathbf{h}_{(t-1)}+\mathbf{b}_{f}\right)
		\end{equation}
		\begin{equation}
			\mathbf{o}_{(t)}=\sigma\left(\mathbf{W}_{x o}^{T} \cdot \mathbf{x}_{(t)}+\mathbf{W}_{h o}^{T} \cdot \mathbf{h}_{(t-1)}+\mathbf{b}_{o}\right)
		\end{equation}
		\begin{equation}
			\mathbf{g}_{(t)}=\tanh \left(\mathbf{W}_{x g}^{T} \cdot \mathbf{x}_{(t)}+\mathbf{W}_{h g}^{T} \cdot \mathbf{h}_{(t-1)}+\mathbf{b}_{g}\right)
		\end{equation}
		\begin{equation}
			\mathbf{c}_{(t)}=\mathbf{f}_{(t)} \otimes \mathbf{c}_{(t-1)}+\mathbf{i}_{(t)} \otimes \mathbf{g}_{(t)}
		\end{equation}
		\begin{equation}
			\label{LSTM-6}
			\mathbf{y}_{(t)}=\mathbf{h}_{(t)}=\mathbf{o}_{(t)} \otimes \tanh \left(\mathbf{c}_{(t)}\right)
		\end{equation}
		As illustrated by \reffig{BiLSTM-with-Attention}, three kinds of gates manipulate and control the memories of EEG signals, namely, the input gate, forget gate, and output gate. Demonstrated by the $\mathbf{i}_{(t)}$, the input gate partially stores the information of $\mathbf{x}_{(t)}$, and controls which part of it should be added to the long-term state $\mathbf{c}_{(t)}$. The forget gate that is controlled by the $\mathbf{f}_{(t)}$ decides which piece of the $\mathbf{c}_{(t)}$ should be overlooked. And the output gate, controlled by $\mathbf{o}_{(t)}$, allows which part of the info from $\mathbf{c}_{(t)}$ should output, denoted as $\mathbf{y}_{(t)}$, as known as the short-term state $\mathbf{h}_{(t)}$. Manipulated by the above gates, two kinds of states are stored. The long-term state $\mathbf{c}_{(t)}$ travels through the cell from left to right, dropping some memories at the forget gate and adding something new from the input gate. After that, the info passes through a non-linear activation function, tanh activation function usually, and then it is filtered by the output gate. In such a way, the short-term state $\mathbf{h}_{(t)}$ is produced. 
		
		\refequ{LSTM-1}$-$\refequ{LSTM-6} describe the procedure of an LSTM cell, where $\mathbf{W}$ and $\mathbf{b}$ are the weights and biases for different layers to store the memory and learn a generalized model, and $\sigma$ is a non-linear activation function, i.e., sigmoid function used in the experiments. For bidirectional LSTM, BiLSTM for short, the signals $\mathbf{x}_{(t)}$ are inputted from left to right for the forward LSTM cell. What is more, they are reversed and inputted into another LSTM cell, the backward LSTM. Thus it leaves us two output vectors, which store much more comprehensive information than a single LSTM cell. Then, they are concatenated as the final output of the cell.
		
		\subsubsection{Attention Mechanism}\label{Attention Mechanism}
		The Attention mechanism, imitated from the human visual, has a vital part to play in the field of Computer Vision (CV), Natural Language Processing (NLP), and Automatic Speech Recognition (ASR)\cite{bahdanau2014neural, xu2015show, yang2016hierarchical, chorowski2015attention}. Not all the signals contribute equally towards the classification. Hence, an Attention mechanism $\mathbf{s}_{(t)}$ is jointly trained as a weighted sum of the output of the BiLSTM with Attention based on the weights. 
		\begin{equation}
			\mathbf{u}_{(t)}=\tanh \left(\mathbf{W}_{w} \mathbf{y}_{(t)}+\mathbf{b}_{w}\right)
		\end{equation}
		\begin{equation}
			\mathbf{\alpha}_{(t)}=\frac{\exp \left(\mathbf{u}_{(t)}^{\top} \mathbf{u}_{w}\right)}{\sum_{t} \exp \left(\mathbf{u}_{(t)}^{\top} \mathbf{u}_{w}\right)}
		\end{equation}
		\begin{equation}
			\mathbf{s}_{(t)}=\sum_{t} \mathbf{\alpha}_{(t)} \mathbf{y}_{(t)}
		\end{equation}
		$\mathbf{u}_{(t)}$ is a Fully-connected (FC) layer for learning features of $\mathbf{y}_{(t)}$, followed by a Softmax layer which outputs a probability distribution $\mathbf{\alpha}_{(t)}$. $\mathbf{W}_{w}$, $\mathbf{u}_{w}$, and $\mathbf{b}_{w}$ denote trainable weights and bias, respectively. It selects and extracts the most significant temporal and spatial information from $\mathbf{y}_{(t)}$ by multiplying $\mathbf{\alpha}_{(t)}$ with regard to the contribution to the decoding tasks. 
 		\begin{figure*}[h]
 		\centering
 		\begin{minipage}[t]{.245\linewidth}
 			\includegraphics[width=1.85in]{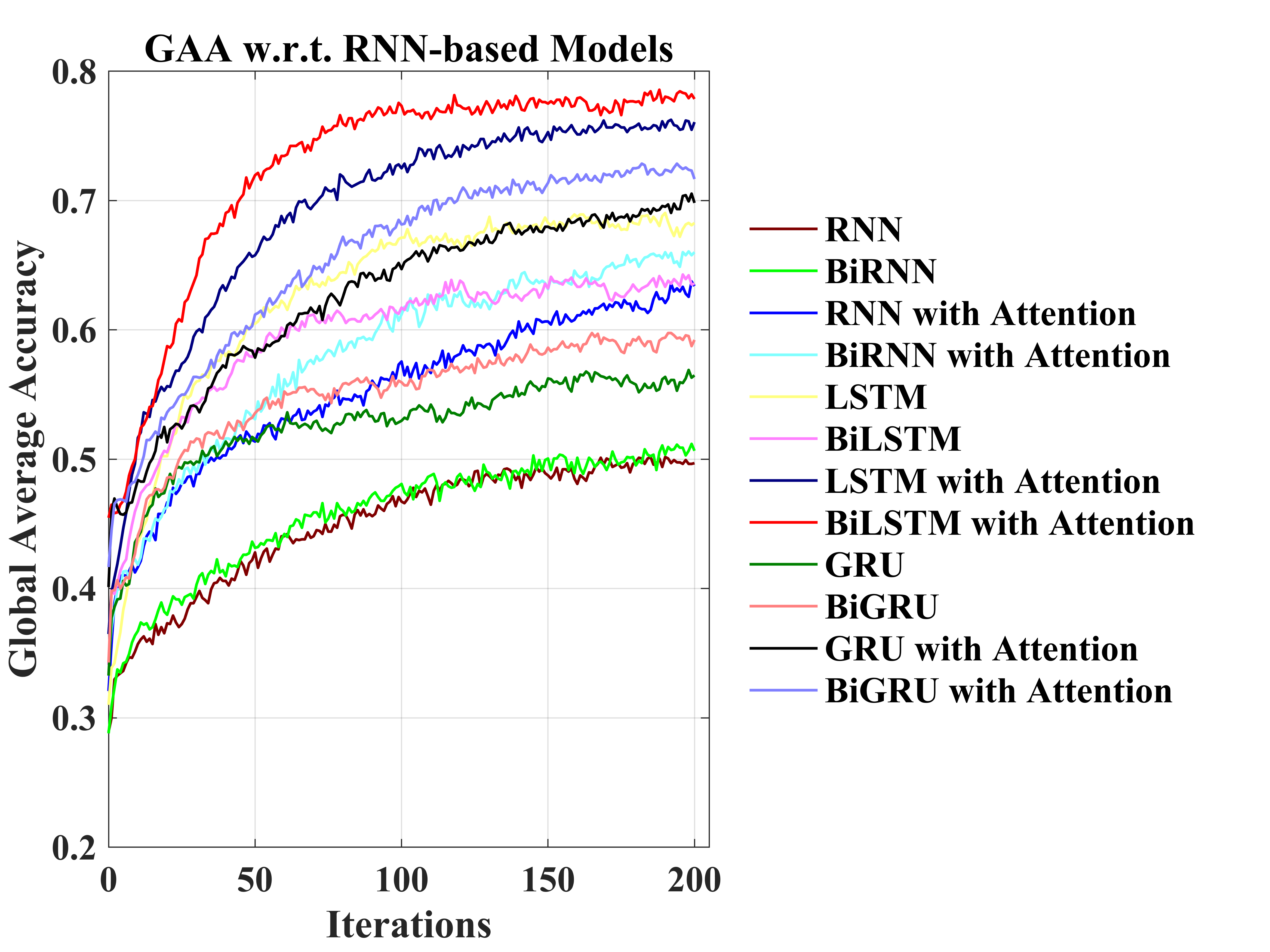}
 			\subcaption{GAA w.r.t. RNN-based Models}
 			\label{GAA_RNN_basedModels}
 		\end{minipage}
 		\begin{minipage}[t]{.245\linewidth}
 			\includegraphics[width=1.85in]{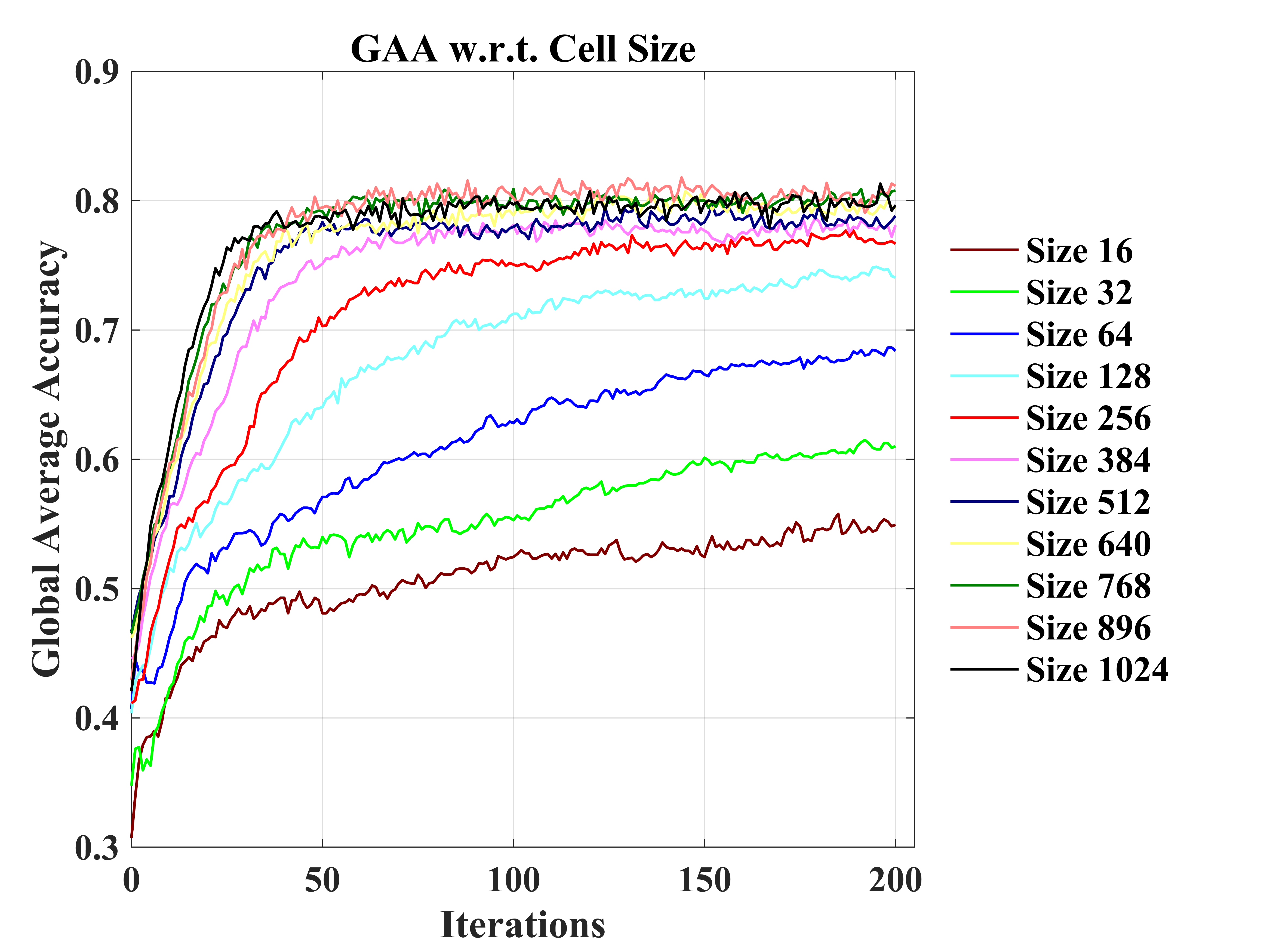}
 			\subcaption{GAA w.r.t. the BiLSTM Cell Size}
 			\label{GAA_Cell_Size}
 		\end{minipage}
 		\begin{minipage}[t]{.245\linewidth}
 			\includegraphics[width=1.85in]{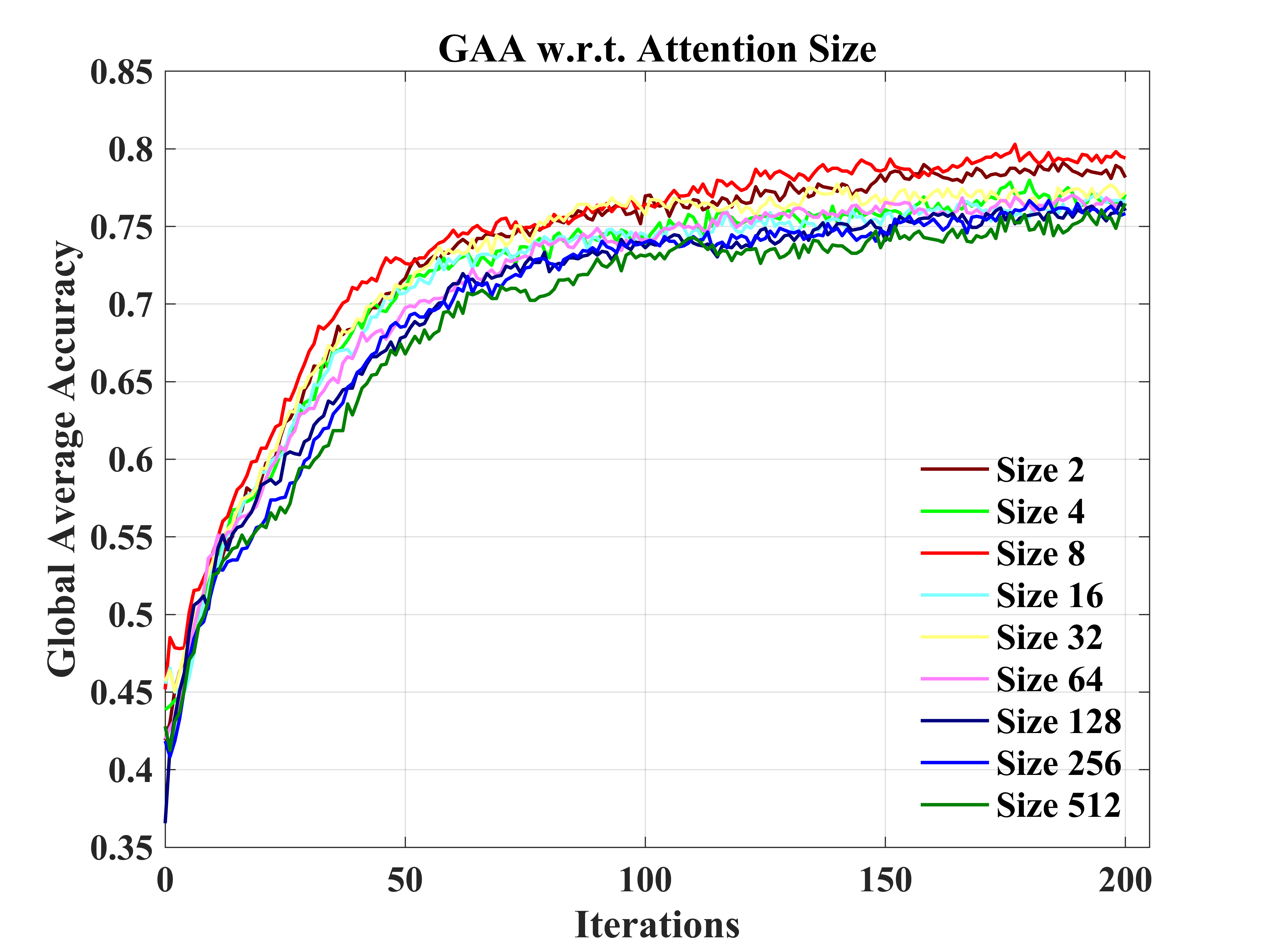}
 			\subcaption{GAA w.r.t. the Attention Size of the BiLSTM}
 			\label{GAA_Attention_Size}
 		\end{minipage}
 		\begin{minipage}[t]{.245\linewidth}
 			\includegraphics[width=1.85in]{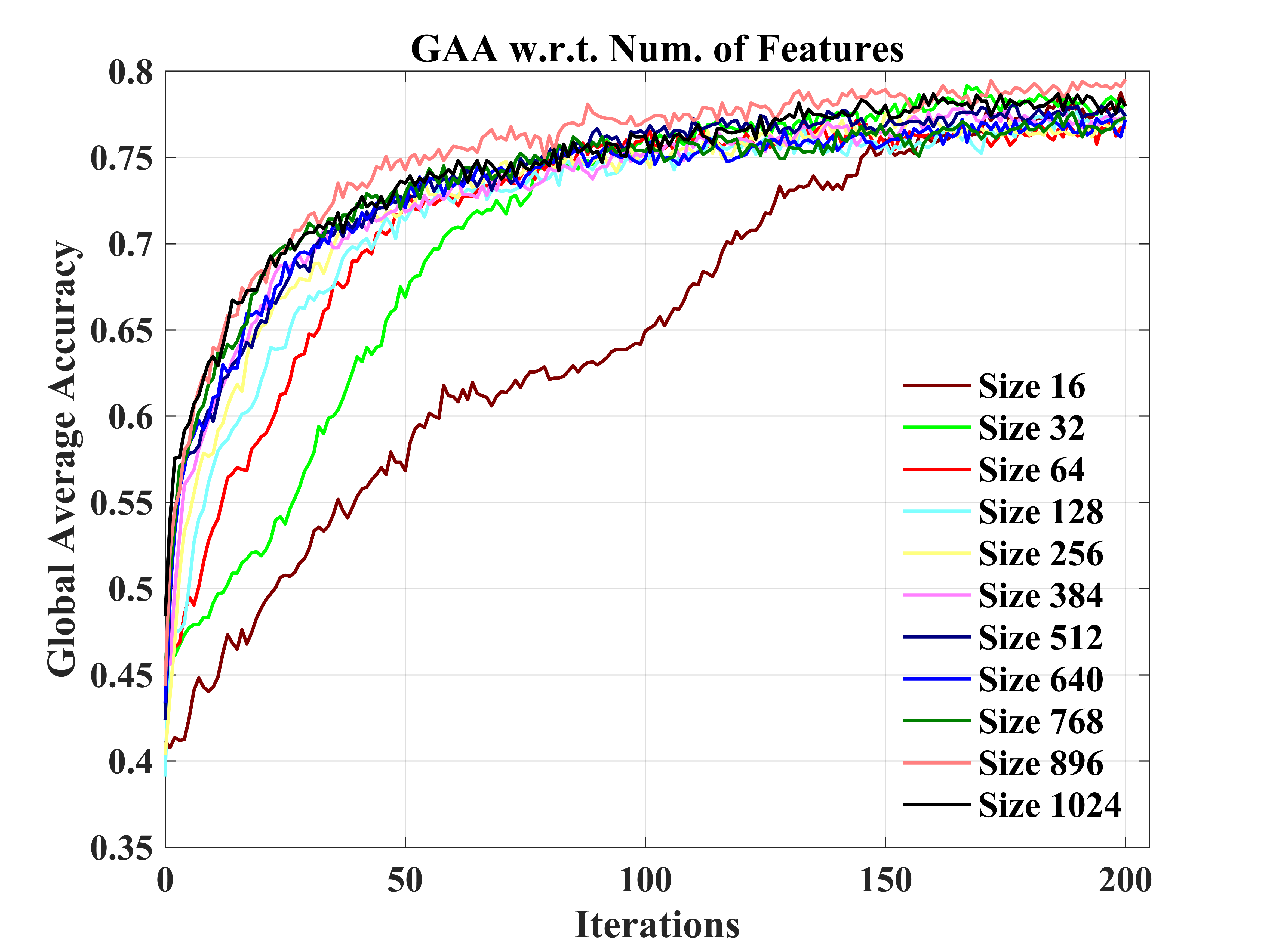}
 			\subcaption{GAA w.r.t. Num. of Extracted Features}
 			\label{GAA_Num_Features}
 		\end{minipage}
 		\begin{minipage}[t]{.245\linewidth}
 			\includegraphics[width=1.85in]{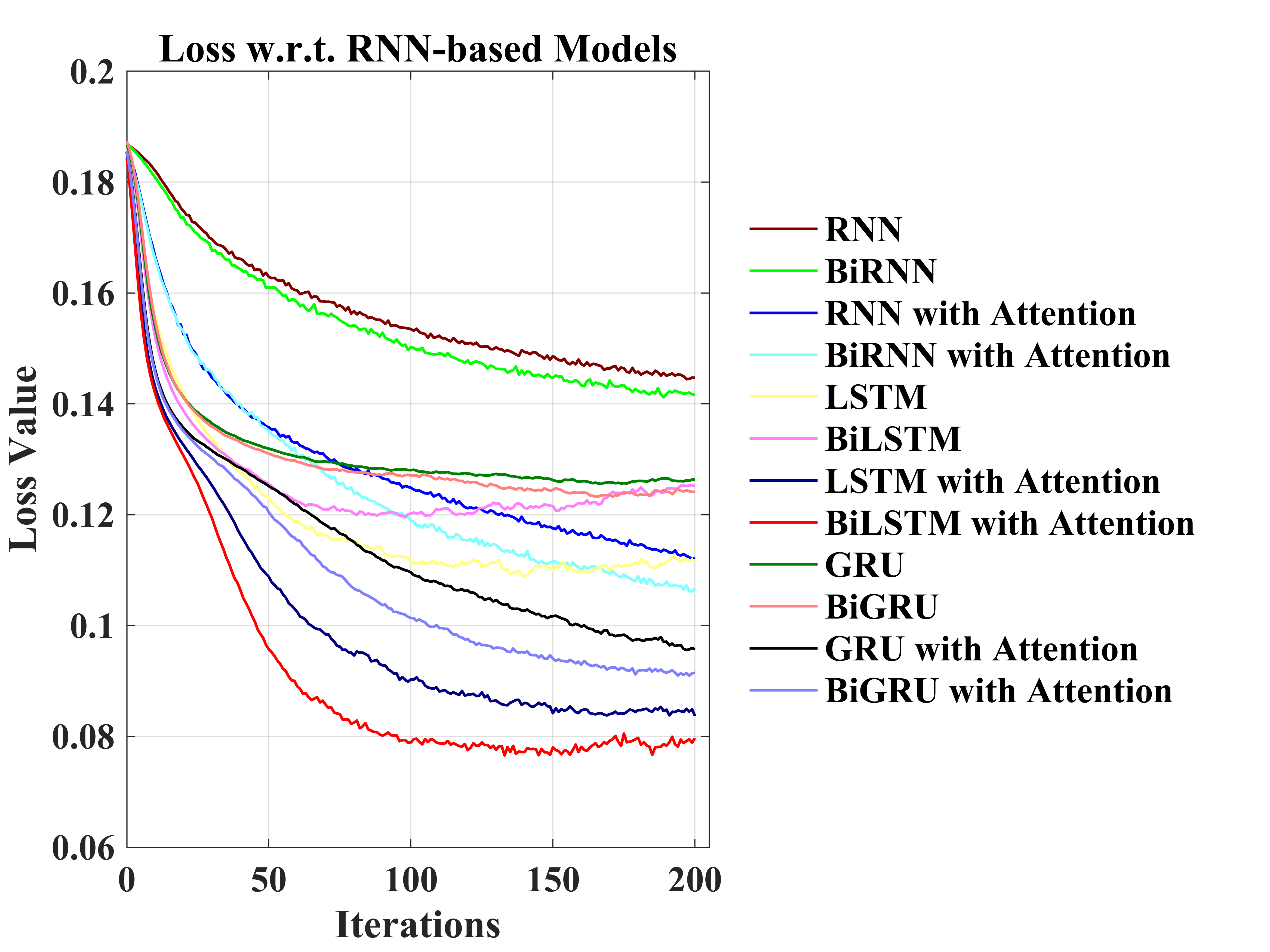}
 			\subcaption{Loss w.r.t. RNN-based Models}
 			\label{Loss_RNN_basedModels}
 		\end{minipage}
 		\begin{minipage}[t]{.245\linewidth}
 			\includegraphics[width=1.85in]{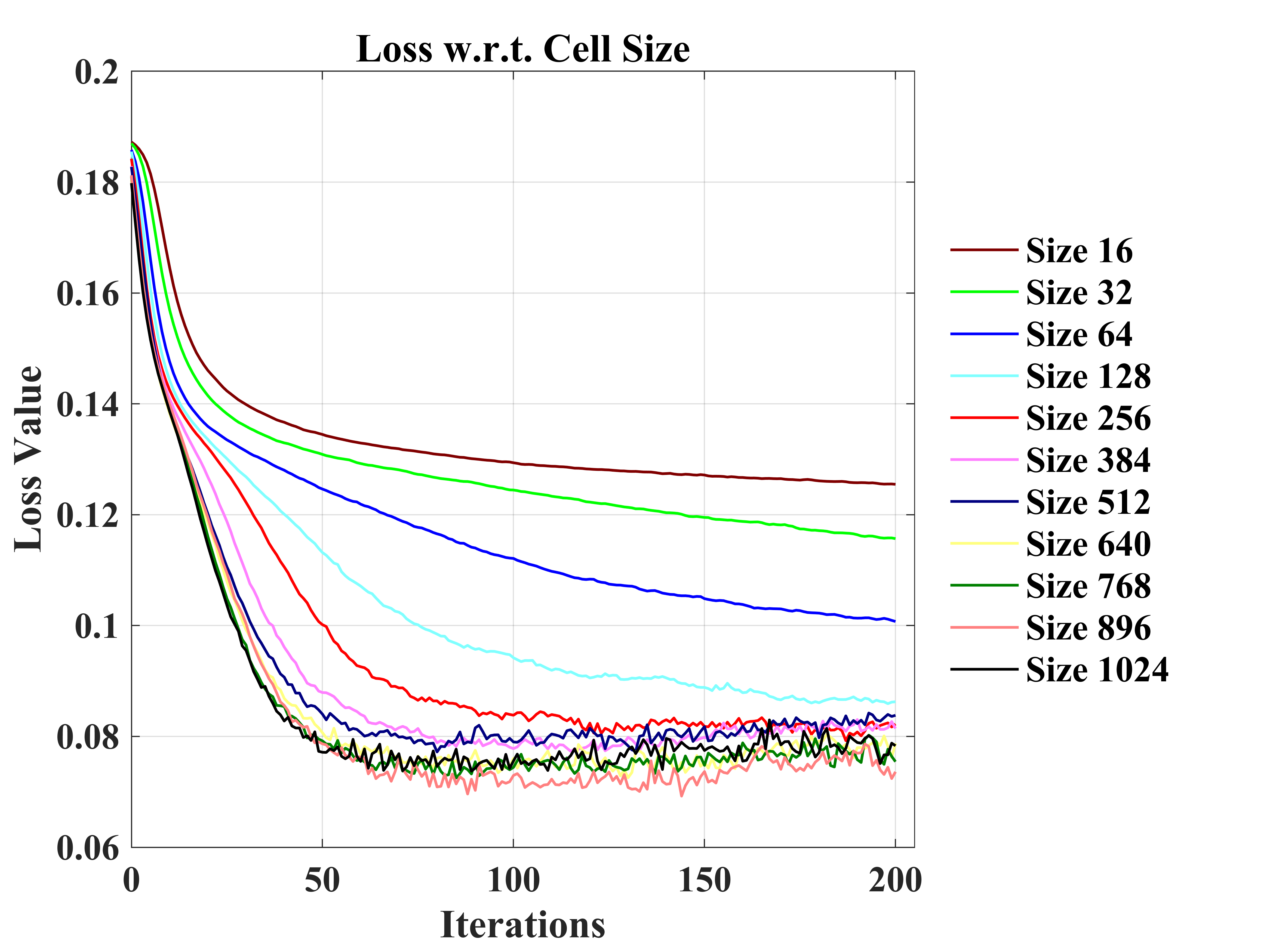}
 			\subcaption{Loss w.r.t. the BiLSTM Cell Size}
 			\label{Loss_Cell_Size}
 		\end{minipage}
 		\begin{minipage}[t]{.245\linewidth}
 			\includegraphics[width=1.85in]{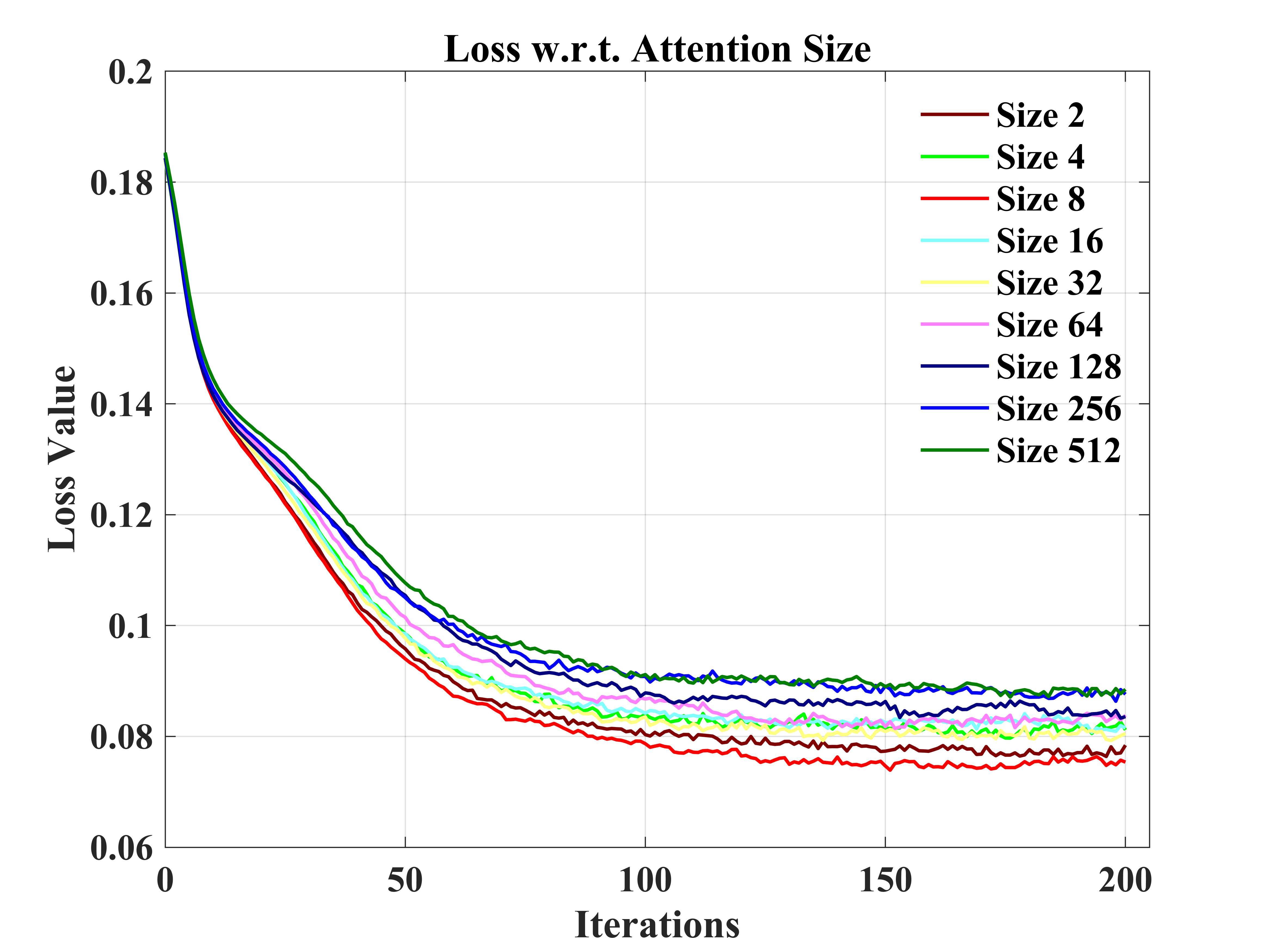}
 			\subcaption{Loss w.r.t. the Attention Size of the BiLSTM}
 			\label{Loss_Attention_Size}
 		\end{minipage}
 		\begin{minipage}[t]{.245\linewidth}
 			\includegraphics[width=1.85in]{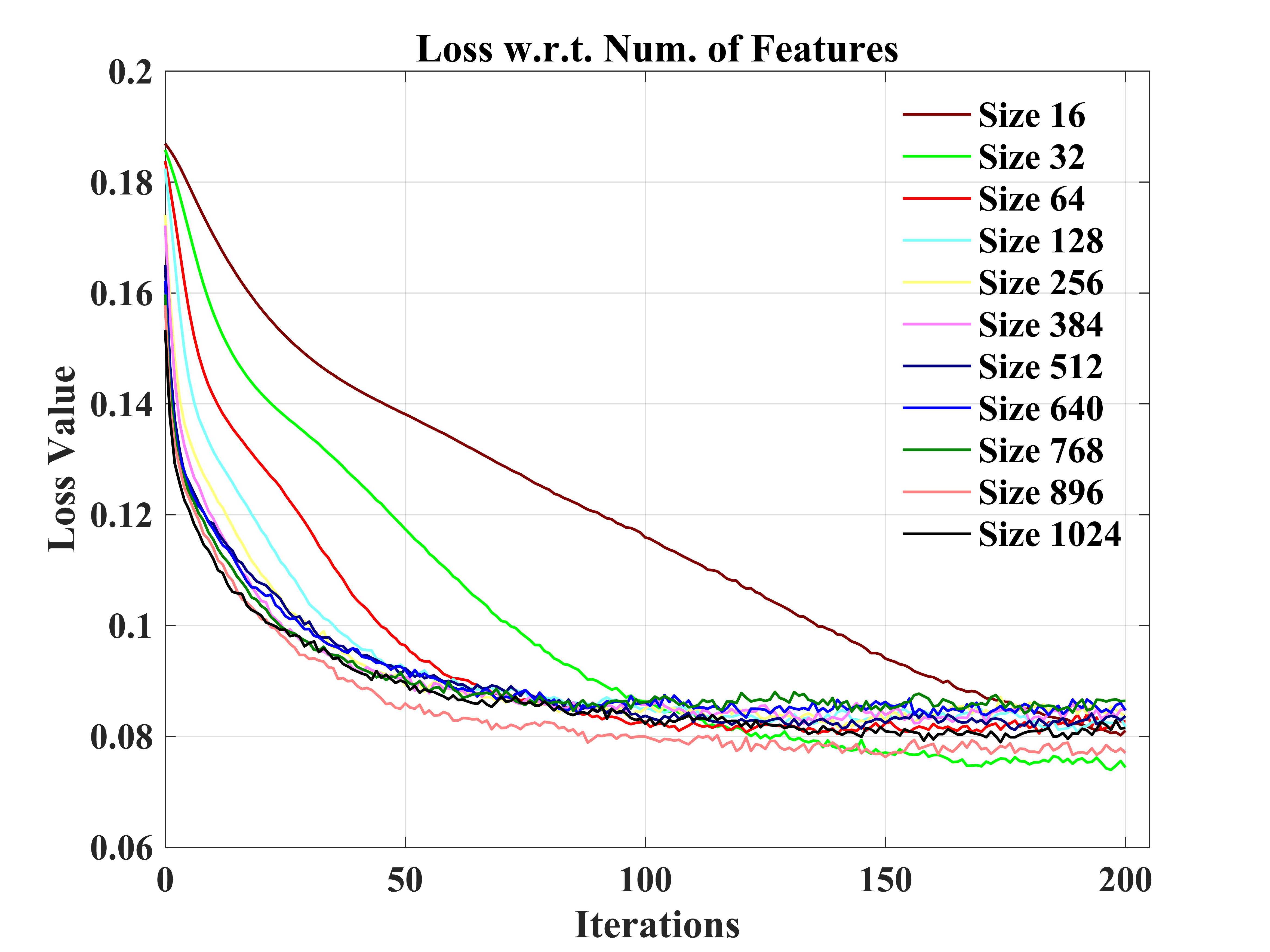}
 			\subcaption{Loss w.r.t. Num. of Extracted Features}
 			\label{Loss_Num_Features}
 		\end{minipage}
 		\caption{Models and Hyperparameters Comparison w.r.t. the RNN-based Methods for Feature Extraction}
 		\label{BiLSTM_With_Attention}
 	\end{figure*}
 
	\subsection{Graph Convolutional Neural Network}\label{Graph Convolutional Neural Network}
		\subsubsection{Graph Convolution}\label{Graph Convolution}
		\begin{figure}[h]
			\centering
			\begin{minipage}[t]{.48\linewidth}
				\includegraphics[width=1.8in]{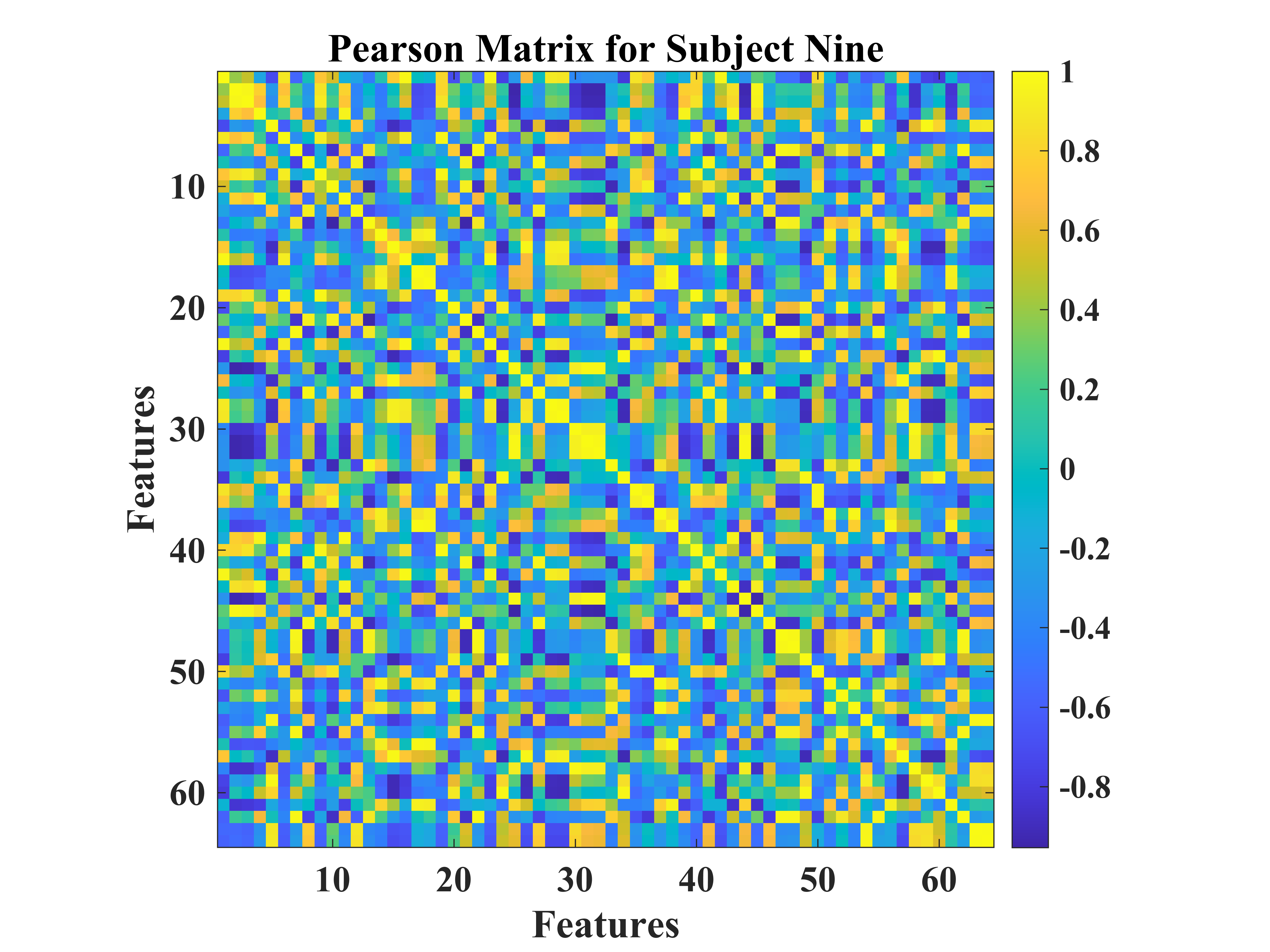}
				\subcaption{Pearson Matrix for Subject Nine}
				\label{Pearson_matrix_for_Subject_Nine}
			\end{minipage}
			\begin{minipage}[t]{.48\linewidth}
				\includegraphics[width=1.8in]{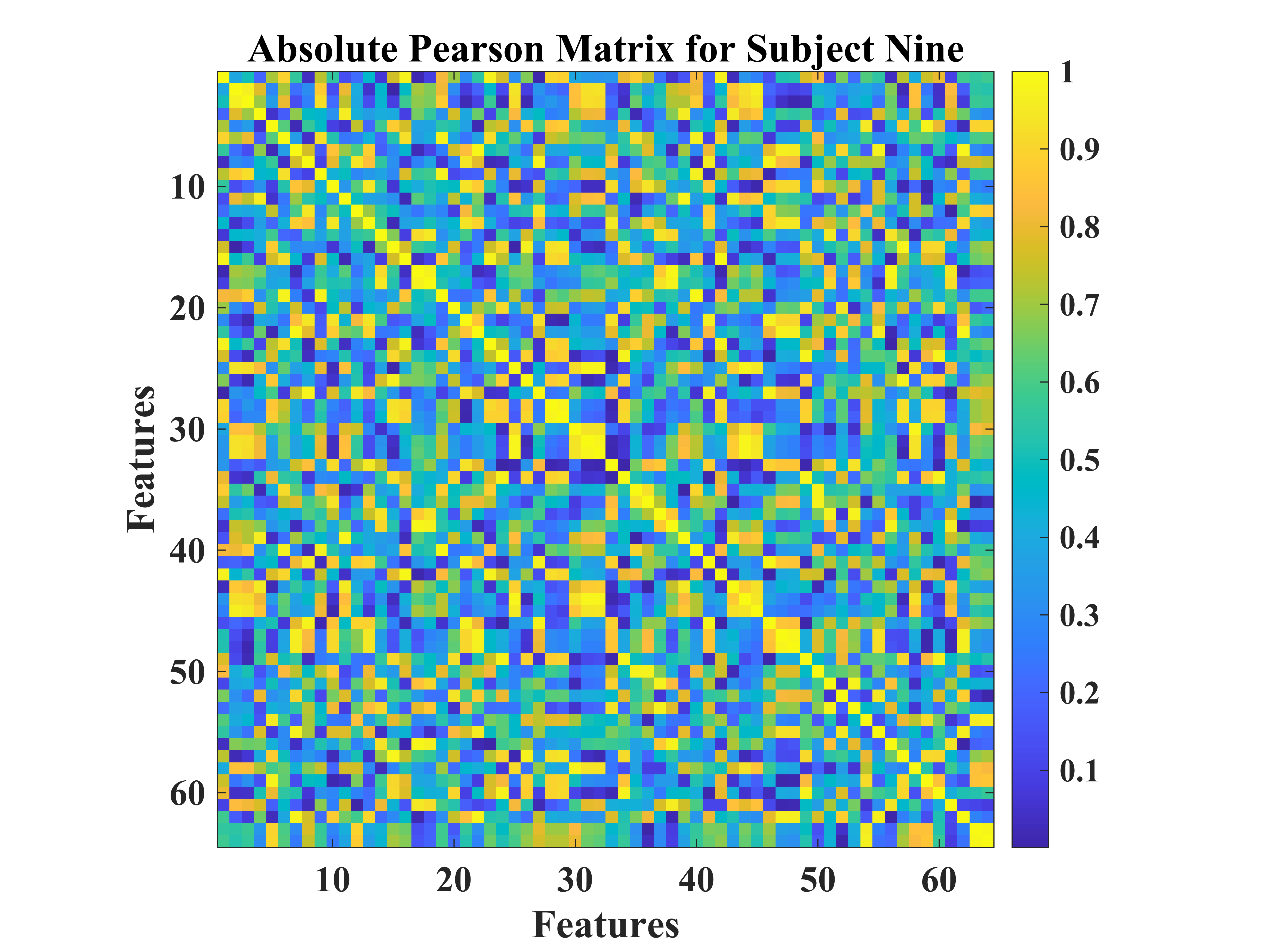}
				\subcaption{Absolute Pearson Matrix for Subject Nine}
				\label{Absolute_Pearson_matrix_for_Subject_Nine}
			\end{minipage}
			\begin{minipage}[t]{.48\linewidth}
				\includegraphics[width=1.8in]{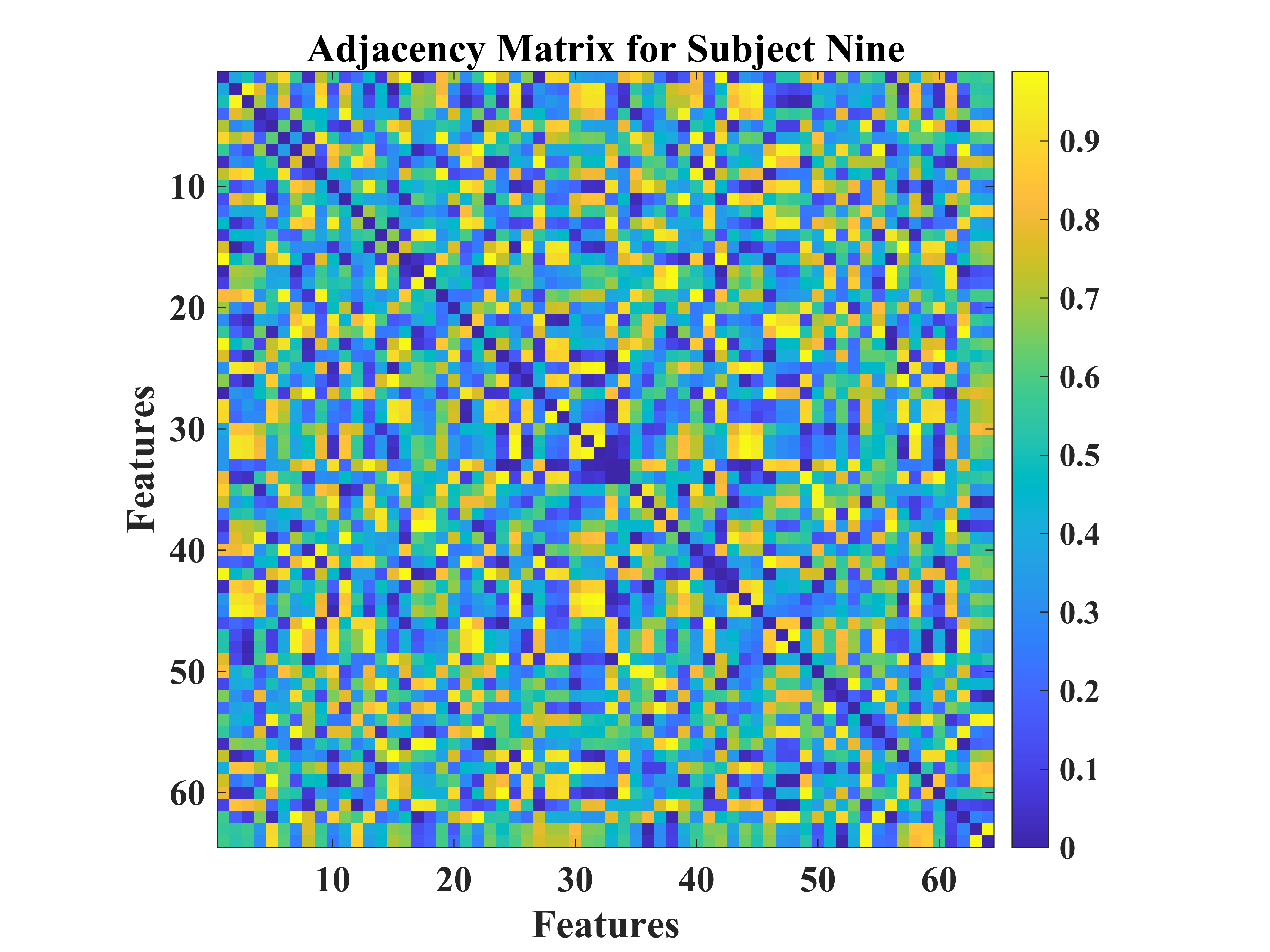}
				\subcaption{Adjacency Matrix for Subject Nine}
				\label{Adjacency_Matrix_for_Subject_Nine}
			\end{minipage}
			\begin{minipage}[t]{.48\linewidth}
				\includegraphics[width=1.8in]{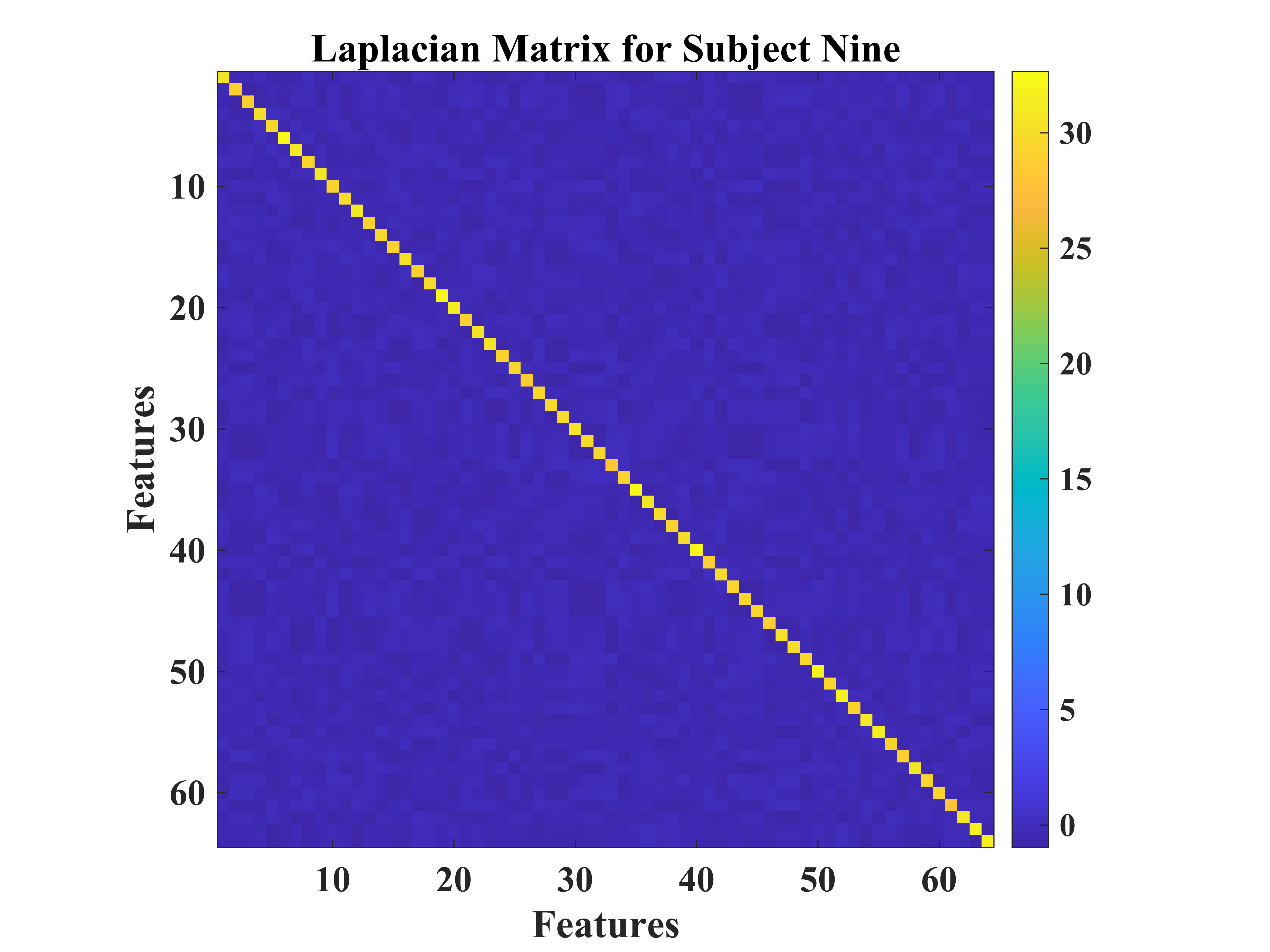}
				\subcaption{Laplacian Matrix for Subject Nine}
				\label{Laplacian_Matrix_for_Subject_Nine}
			\end{minipage}
			\caption{The Pearson, Absolute Pearson, Adjacency, and Laplacian Matrices for Subject Nine.}
			\label{Subject-Nine}
		\end{figure}
		In the graph theory, a graph is presented by the graph Laplacian $L$. It is computed by the Degree Matrix $D$ minus the Adjacency Matrix $A$, i.e., $L=D-A$. In this work, Pearson's matrix $P$ was utilized to measure the inner correlations among features. 
		\begin{equation}
			P_{X, Y}=\frac{E\left(\left(X-\mu_{X}\right)\left(Y-\mu_{Y}\right)\right)}{\sigma_{X} \sigma_{Y}}
		\end{equation}
		where $X$ and $Y$ are two variables regarding different features, $\rho_{X, Y}$ is their correlation, $\sigma_{X}$ and $\sigma_{Y}$ are the standard deviations and $\mu_{X}$ and $\mu_{Y}$ are the expectations. Besides, the Adjacency Matrix $A$ is recognised as:
		\begin{equation}
			A=|P|-I
		\end{equation}
		Where $|P|$ is the absolute of the Pearson's matrix $P$, and $I \in \mathbb{R}\textsuperscript{\emph{N}$\times$\emph{N}}$ is an identity matrix. In addition, the Degree Matrix $D$ of the graph is computed as follows.
		\begin{equation}
			D_{ii}=\sum_{j=1}^{N} A_{ij}
		\end{equation}
		Then, the normalized graph Laplacian is computed as:
		\begin{equation}
			L=D-A=I_{N}-D^{-1 / 2} A D^{-1 / 2}
		\end{equation}
		It is then decomposed by the Fourier basis $U=\left[u_{0}, \ldots, u_{N-1}\right] \in \mathbb{R}^{N \times N}$. The graph Laplacian is described as $L=U\Lambda U^{T}$, where $\Lambda=\operatorname{diag}\left(\left[\lambda_{0}, \ldots, \lambda_{N-1}\right]\right) \in \mathbb{R}^{N \times N}$ are the eigenvalues of $L$. The convolutional operation for a graph is defined as:
		\begin{equation}
			y=g_{\theta}(L) x=g_{\theta}\left(U \Lambda U^{T}\right) x=U g_{\theta}(\Lambda) U^{T} x 
		\end{equation}
		In which, $g_{\theta}$ is a non-parametric filter. Specifically, the operation is as follows.
		\begin{equation}
			y_{:, j}^{k+1}=\sigma\left(\sum_{i=1}^{f_{k-1}} U g_{\theta}(\Lambda) U^{T} x_{:, i}^{k}\right)
		\end{equation}
		In which $y^{k} \in \mathbb{R}^{N \times f_{k-1}}$ denotes the signals, $N$ is the number of vertices of the graph, $f_{k-1}$ and $f_{k}$ are the numbers of input and output channels respectively, and the $\sigma$ denotes a non-linearity activation function. What is more, $g_{\theta}$ is approximated by the Chebyshev polynomials because it is not localized in space and very time-consuming\cite{hammond2011wavelets}. The Chebyshev recurrent polynomial approximation is described as $T_{k}(x)=2 x T_{k-1}(x)-T_{k-2}(x), \ T_{0}=1, \ T_{1}=x$. And the filter can be presented as $g_{\theta}(\Lambda)=\sum_{k=0}^{K-1} \theta_{k} T_{k}(\tilde{\Lambda})$, in which $\theta \in \mathbb{R}^{K}$ is a set of coefficients, and $T_{k}(\tilde{\Lambda}) \in \mathbb{R}^{K}$ is the $k^{th}$ order polynomial at $\tilde{\Lambda}=2 \Lambda / \lambda_{m a x}-I_{n}$, and $I_{n} \in [-1, 1]$ is a diagonal matrix of the scaled eigenvalues. The Convolution can be rewritten as:
		\begin{equation}
			y=\sum_{k=0}^{K-1} \theta_{k} T_{k}(\tilde{L}) x
		\end{equation}
		
		\subsubsection{Graph Pooling}\label{Graph Pooling}
		The graph pooling operation can be achieved via the Graclus multilevel clustering algorithm, which consists of nodes clustering and one-dimensional pooling\cite{dhillon2007weighted}. A greedy algorithm is implemented to compute the successive coarser of a graph and minimize the clustering objective, from which the normalized cut is chosen\cite{shi2000normalized}. Through such a way, meaningful neighborhoods on graphs are acquired. Reference \cite{defferrard2016convolutional} proposed to carry out a balanced binary tree to store the neighborhoods, and a one-dimensional pooling is then applied for precise dimensionality reduction. 

	\subsection{Proposed Approach}\label{Proposed Approach}
	The presented approach was a combination of the Attention-based BiLSTM and the GCN as illustrated in \reffig{Framework}. The BiLSTM with the Attention mechanism was presented to derive relevant features from raw EEG signals. During the procedure, features were obtained from neurons at the FC layer. The GCN was then applied to classify the extracted features. It was the combination of two models that promoted and enhanced the decoding performance by a significant margin compared with existing studies. Details were provided in the following.
	
	First of all, an optimal RNN-based model was explored to obtain relevant features from raw EEG signals. As detailed in \reffig{BiLSTM_With_Attention}, in this work, the BiLSTM with Attention model was best performed, which achieved 77.86\% Global Average Accuracy (GAA). The input size $\mathbf{x}_{(t)}$ of the model was 64 denoting 64 channels (electrodes) of raw EEG signals. The max time $t$ was chosen as 64, which was 0.4 seconds' segment. According to \reffig{GAA_Cell_Size}, a higher accuracy has obtained while increasing the number of cells of BiLSTM model. It should, however, be noted in \reffig{Loss_Cell_Size} that when there were more than 256 cells, the loss showed an upward trend, which indicated the concern of overfitting due to the increment of the model complexity. As a result, 256 LSTM cells (76.67\% GAA) were chosen to generalize the model. Meanwhile, it was apparent that, in \reffig{GAA_Attention_Size}, as for the linear size of the Attention weights, the majority of the choices did not make a difference. Thus, 8 neurons, with 79.40\% GAA, were applied during the experiments empirically. Comparing \reffig{GAA_Num_Features} and \reffig{Loss_Num_Features}, it showed that a compromise solution should be applied, which took consideration of both performance and input size of the GCN. As a result, the linear size of 64 (76.73\% GAA) was utilized at the FC layer.
		
	Besides, to prevent overfitting, a 25\% dropout\cite{srivastava2014dropout} for the BiLSTM and FC layer was implemented. The model carried out batch normalization (BN)\cite{ioffe2015batch} for the FC layer, which was activated by the softplus function\cite{hahnloser2000digital}. And L2 norm with $1\times10^{-7}$ coefficient was applied to the Euclidean distance as the loss function. 1,024 batch sizes were used to maximize the usage of GPU resources. $1\times10^{-4}$ learning rate was applied to Adam Optimizer\cite{kingma2014adam}.
	
	Furthermore, 2$^{nd}$ order Chebyshev polynomial was applied to approximate convolutional filters in the experiments. The GCN consisted of six graph convolutional layers with 16, 32, 64, 128, 256, and 512 filters, respectively, each followed by a graph max-pooling layer, and a softmax layer derived the final prediction.
	
	In addition, for the GCN model, before the non-linear softplus activation function, BN was utilized at all of the layers except the final softmax. $1\times10^{-7}$ L2 norm was added to the loss function, which was a cross-entropy loss. Stochastic Gradient Descent\cite{zhang2004solving} with 16 batch sizes was optimized by the Adam ($1\times10^{-7}$ learning rate). 
	
	All the experiments above were performed and implemented by the Google TensorFlow\cite{abadi2016tensorflow} 1.14.0 under NVIDIA RTX2080ti and CUDA10.0.
	
\section{Results and Discussion}\label{Results and Discussion}
	\subsection{Description of Dataset}\label{Description of Dataset}
	The data collected from the EEG Motor Movement/Imagery Dataset\cite{goldberger2000physiobank} was employed in this study. Numerous EEG trials were acquired from 109 participants performing four MI tasks, i.e., imagining left fist (L), right fist (R), both fists (B), and both feet (F) (21 trials per task). Each trial is a 4-second experiment's duration (160 Hz sample rate) with one single task\cite{hou2019novel}. In this work, a 0.4 seconds' temporal segment of 64 channels' signals, i.e., 64 channels $\times$ 64 time points, was regarded as a sample. In \refsec{Group-wise Prediction}, we used a group of 20 subjects' data ($S_1-S_{20}$) to train and validate our method. 10-fold cross validation was carried out. Further, 50 subjects ($S_1-S_{50}$) were selected to verify the repeatability and stability of our approach. In \refsec{Subject-specific Adaptation}, the data set of individual subjects ($S_1-S_{10}$), was utilized to perform subject-level adaptation. For all the experiments, the dataset was randomly divided into 10 parts, where 90\% was the training set, and the left 10\% was regarded as the test set. In \refsec{Group-wise Prediction}, the above procedure has carried out for 10 times. Thus, it left us 10 results of 10-fold cross validation.
	
	\subsection{Group-wise Prediction}\label{Group-wise Prediction}
\begin{figure}[h]
		\centering
		\begin{minipage}[t]{.48\linewidth}
			\includegraphics[width=1.8in]{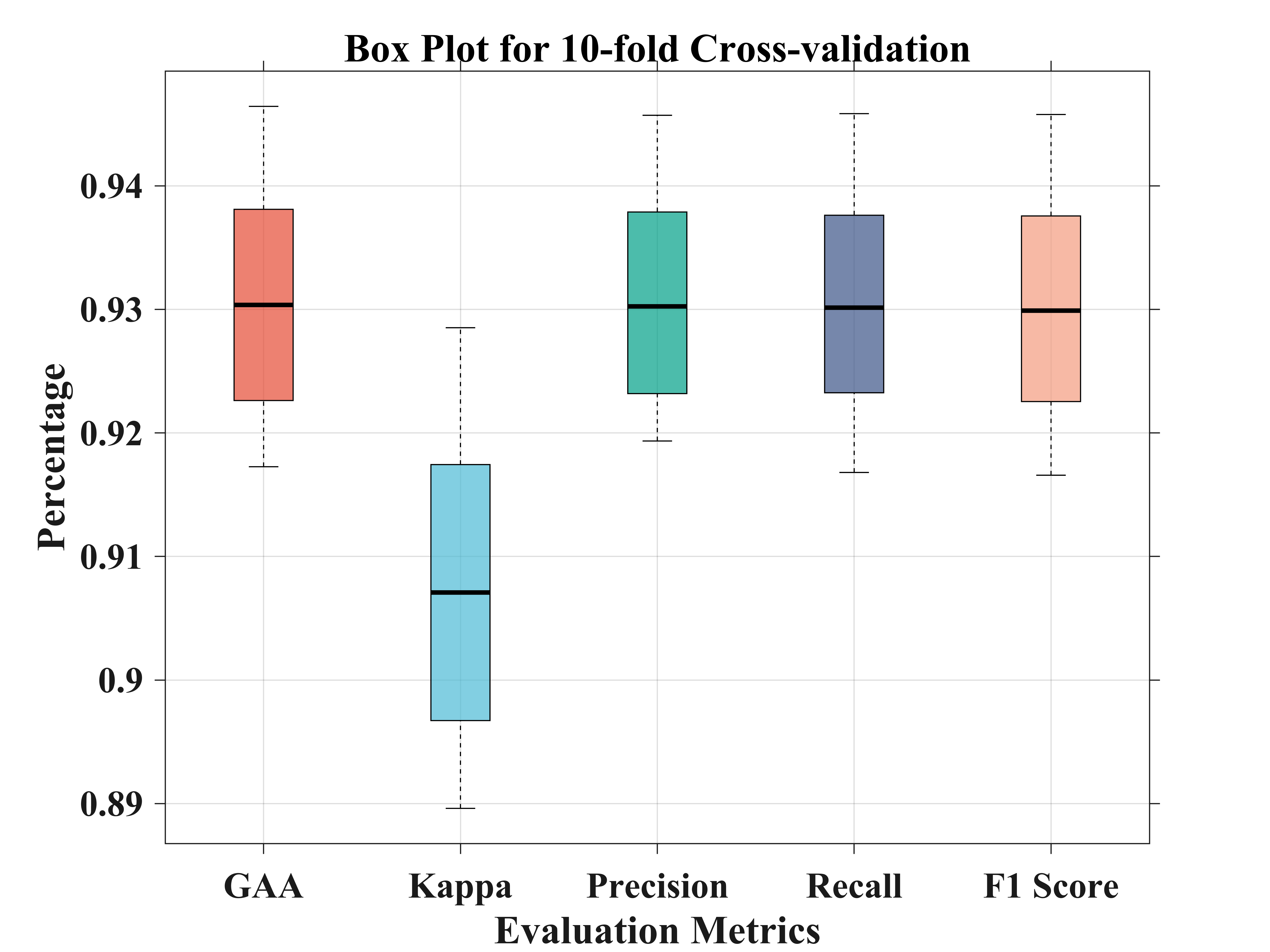}
			\subcaption{Box Plot for 10-fold cross validation}
			\label{box_cross_validation}
		\end{minipage}
		\begin{minipage}[t]{.48\linewidth}
			\includegraphics[width=1.6in]{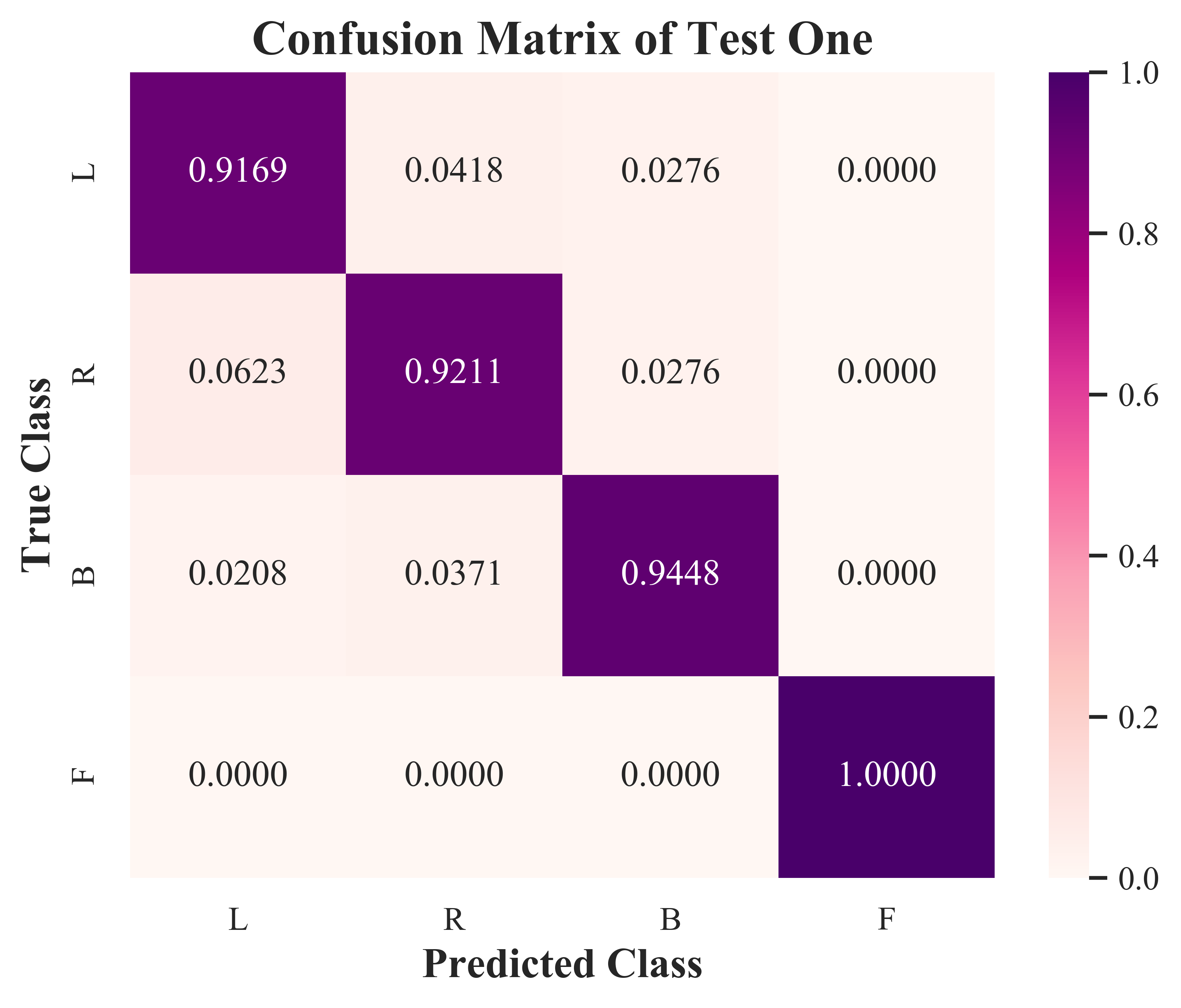}
			\subcaption{Confusion Matrix of Test One}
			\label{Confusion_Matrix}
		\end{minipage}
		\caption{Box plot and confusion matrix for 10-fold cross validation.}
		\label{Loss_Box_Cross_Validation}
	\end{figure}

	It was suggested that the inter-subject variability remains one of the concerns for interpreting EEG signals\cite{tanaka2020group}. Firstly, a small group size (20 subjects) was adopted for group-wise prediction. In \reffig{GAA_RNN_basedModels}, 63.57\% GAA was achieved by the BiLSTM model. After applying the attention mechanism, it enhanced the decoding performance, which accomplished 77.86\% GAA (14.29\% improvement). Further, we employed Attention-based BiLSTM-GCN model in this work. It attained 94.64\% maximum GAA\cite{hou2019novel} (31.07\% improvement compared with the BiLSTM model), and 93.04\% median accuracy from 10-fold cross-validation. Our method promoted the classification capability under subjects' variability and variations by taking the topological relationship of relevant features into consideration. Meanwhile, as illustrated by \reffig{box_cross_validation}, the median values of GAA, Kappa, precision, recall, and F1 Score were 93.04\%, 90.71\%, 93.02\%, 93.01\%, and 92.99\%, respectively. To the knowledge of the authors, the proposed method has achieved the best state-of-the-art performance in group-level prediction. Besides, remarkable results of 10-fold cross-validation have verified the repeatability and stability. Furthermore, the confusion matrix of Test One (94.64\% GAA) was given in \reffig{Confusion_Matrix}. 91.69\%, 92.11\%, 94.48\%, and 100\% accuracy were obtained for each task. It can be observed that our method was unprecedentedly effective and efficient in detecting human motion intents from raw EEG signals.
	
\begin{figure}[h]
		\centering
		\begin{minipage}[t]{.48\linewidth}
			\includegraphics[width=1.8in]{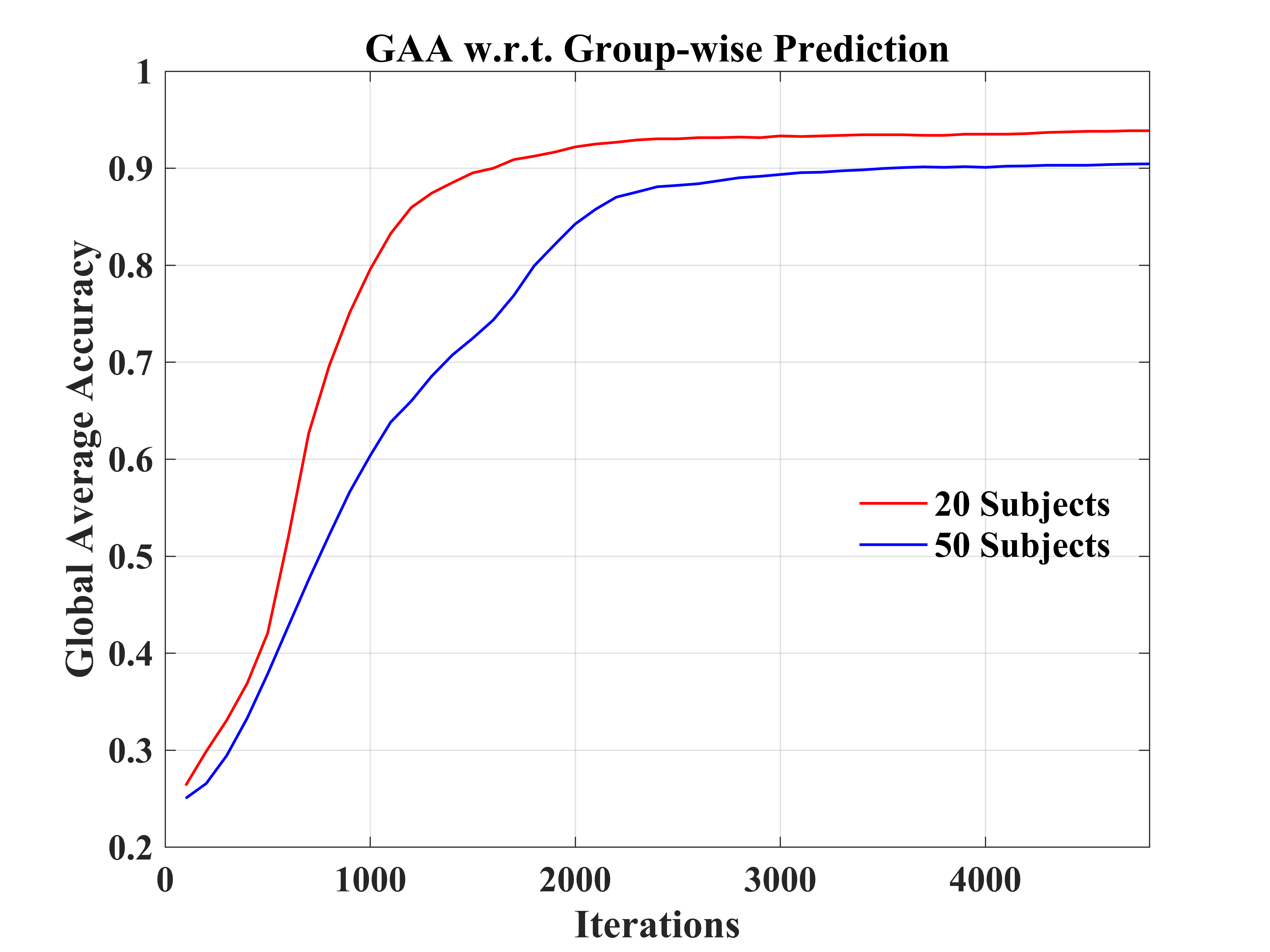}
			\subcaption{GAA w.r.t. group-wise prediction}
			\label{GAA_Group_wise_Prediction}
		\end{minipage}
		\begin{minipage}[t]{.48\linewidth}
			\includegraphics[width=1.65in]{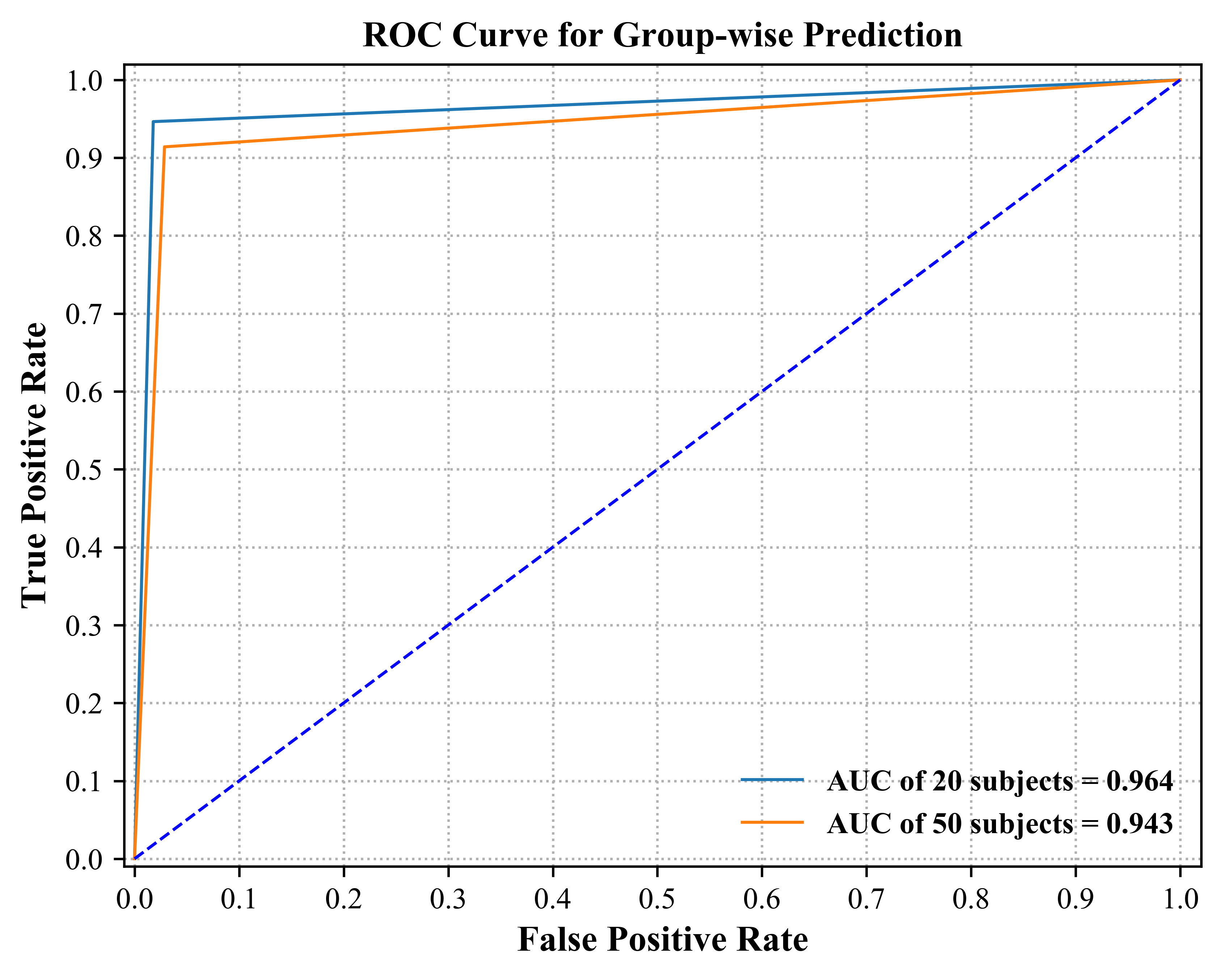}
			\subcaption{ROC Curve w.r.t. group-wise prediction}
			\label{ROC_Group_Wise}
		\end{minipage}
		\caption{GAA and ROC Curve for 20 and 50 subjects, separately.}
		\label{Group_Wise_Prediction}
	\end{figure}

	 By grouping signals from additional 30 subjects (in total 50 subjects), the robustness of the method has been validated in \reffig{Group_Wise_Prediction}. 
	 
	Towards practical EEG-based BCI applications, it is essential to develop a robust model to counter serious individual variability\cite{tanaka2020group}. \reffig{GAA_Group_wise_Prediction} illustrated the GAA of our method through iterations. As listed in \reffig{ROC_Group_Wise}, we can see that 94.64\% and 91.40\% GAA were obtained with regard to the group of 20 and 50 subjects, respectively. And the Area under the Curves (AUC) were 0.964 and 0.943. Indicated by the above results, the presented approach can successfully filter the distinctions of signals, even though the data set was extended. In other words, by increasing the inter-subject variability, the robustness and effectiveness of the method were evaluated.
	
	The comparison of group-wise evaluation was demonstrated measured by the maximum of GAA\cite{hou2019novel} during experiments\cite{ma2018improving, hou2019novel}. Here, we compared the performance of several state-of-the-art methods in \reftab{group-wise-evaluation}.
	
	\begin{table}[H]
		\centering
		\footnotesize
		\caption{Comparison on group-wise evaluation}
		\label{group-wise-evaluation}
		\resizebox{\linewidth}{!}{
			\begin{tabular}{ccccc}
				\hline
				\textbf{Related Work} & \textbf{Max. GAA} & \textbf{Approach} & \tabincell{c}{\textbf{Num of}\\ \textbf{Subjects}} & \textbf{Database} \\ \hline
				Ma \emph{et al.} (2018) & 68.20\% & RNNs & 12 & \multirow{4}{*}{Physionet Database} \\
				\multirow{2}{*}{Hou \emph{et al.} (2019)} & 94.50\% & \multirow{2}{*}{ESI + CNNs} & 10 & \\
				 & 92.50\% & & 14 & \\
				\textbf{This work} & \textbf{94.64\%} & \textbf{Attention-based BiLSTM-GCN} & \textbf{20} & \\ \hline
			\end{tabular}
		}
	\end{table}
	
	\reftab{group-wise-evaluation} listed the performance of related methods. Hou \emph{et al.} achieved competitive results. However, our method obtained higher performance (0.14\% accuracy improvement) even with doubling the number of subjects. It can be found that our approach has outperformed those by giving the highest accuracy of decoding EEG MI signals. 

	\subsection{Subject-specific Adaptation}\label{Subject-specific Adaptation}
	The performance of individual adaptation has witnessed a flourishing increment\cite{ortiz2019new, sadiq2019motor, taran2019motor, zhang2019novel, ji2019eeg, amin2019multilevel, dose2018end, hou2019novel}. The results of our method on subject-level adaptation have been reviewed in \reftab{Subject-level Evaluation}, and we compared the results in \reftab{Subject-level-evaluation}.
	
	\begin{table}[h]
		\caption{Subject-level Evaluation}
		\label{Subject-level Evaluation}
		\resizebox{\linewidth}{!}{
			\begin{tabular}{cccccc}
				\hline
				\textbf{No. of Subject} & \textbf{GAA} & \textbf{Kappa} & \textbf{Precision} & \textbf{Recall} & \textbf{F1 Score} \\ \hline
				1              & 94.05\% & 92.06\% & 94.20\%   & 94.32\% & 94.16\%  \\ 
				2              & 96.43\% & 95.19\% & 96.06\%   & 96.06\% & 96.06\%  \\ 
				3              & 97.62\% & 96.79\% & 97.33\%   & 97.08\% & 97.18\%  \\ 
				4              & 90.48\% & 87.34\% & 91.30\%   & 91.11\% & 90.42\%  \\ 
				5              & 95.24\% & 93.61\% & 95.96\%   & 95.06\% & 95.38\%  \\ 
				6              & 94.05\% & 92.02\% & 93.40\%   & 94.96\% & 93.66\%  \\ 
				7              & 98.81\% & 98.40\% & 98.81\%   & 99.07\% & 98.92\%  \\ 
				8              & 95.24\% & 93.60\% & 95.39\%   & 95.04\% & 95.19\%  \\ 
				9              & 98.81\% & 98.39\% & 99.11\%   & 98.68\% & 98.87\%  \\ 
				10            & 94.05\% & 91.98\% & 93.39\%   & 94.70\% & 93.61\%  \\ 
				\textbf{Average} & \textbf{95.48\%} & \textbf{93.94\%} & \textbf{95.50\% } & \textbf{95.61\%} & \textbf{95.35\%} \\ \hline
			\end{tabular}
		}
	\end{table}

	Results were given in \reftab{Subject-level Evaluation}, from which the highest GAA was 98.81\% achieved by the Subject $S_7$ and $S_9$, and the lowest was 90.48\% by $S_4$. On average, the presented approach can handle the challenge of subject-specific adaptation. It achieved competitive results, with an average accuracy of 95.48\%. Moreover, Cohen's Kappa coefficient (Kappa), precision, recall, and f1-score were 93.94\%, 95.50\%, 95.61\%, and 95.35\%, respectively. The promising results above indicated that the introduced method filtered raw EEG signals, and succeed in classifying MI tasks.
	
	\begin{figure}[h]
		\centering
		\begin{minipage}[t]{.48\linewidth}
			\includegraphics[width=1.8in]{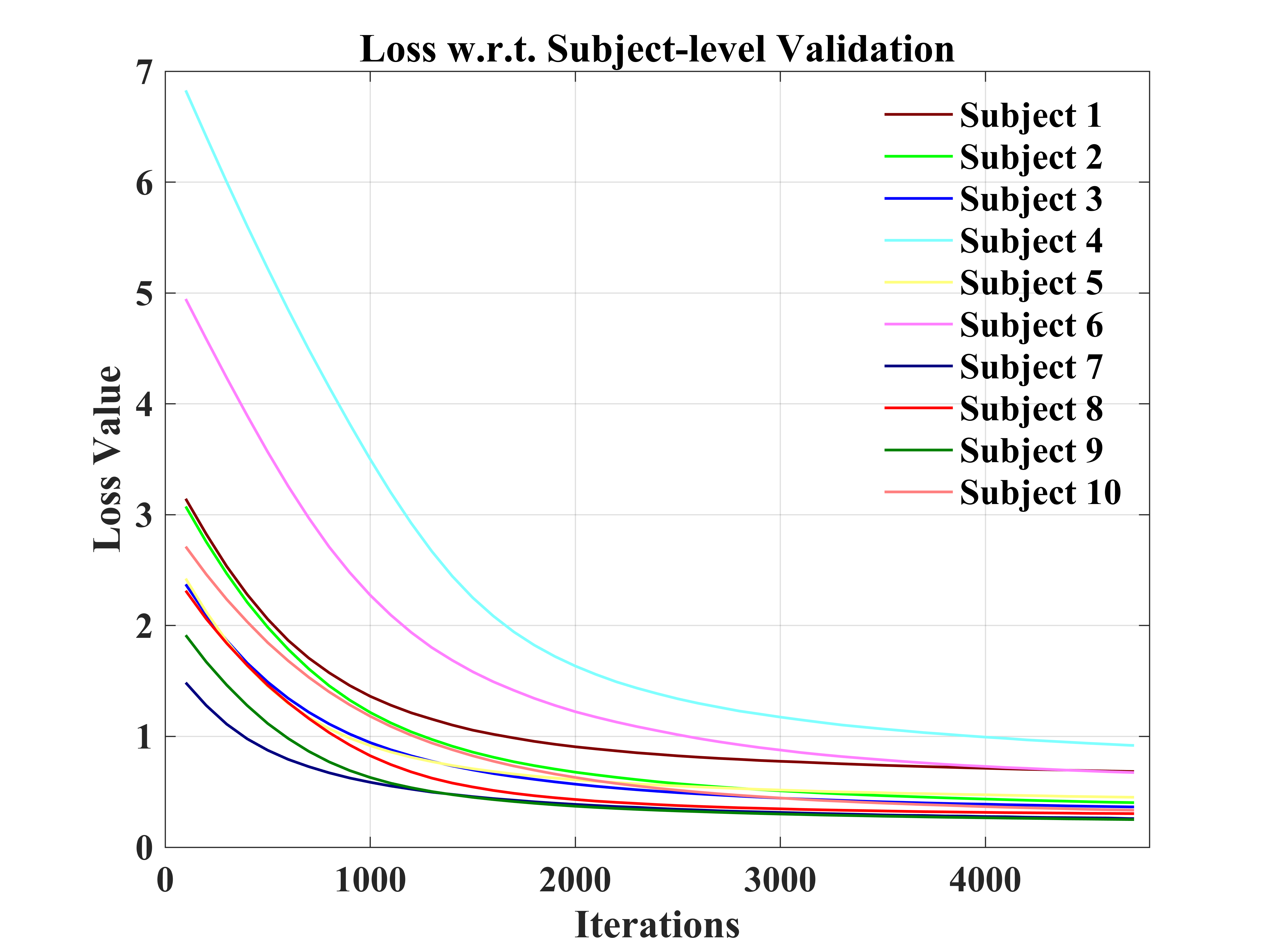}
			\subcaption{Loss w.r.t. Subject-level Validation}
			\label{Loss_Subject_level_Validation}
		\end{minipage}
		\begin{minipage}[t]{.48\linewidth}
			\includegraphics[width=1.65in]{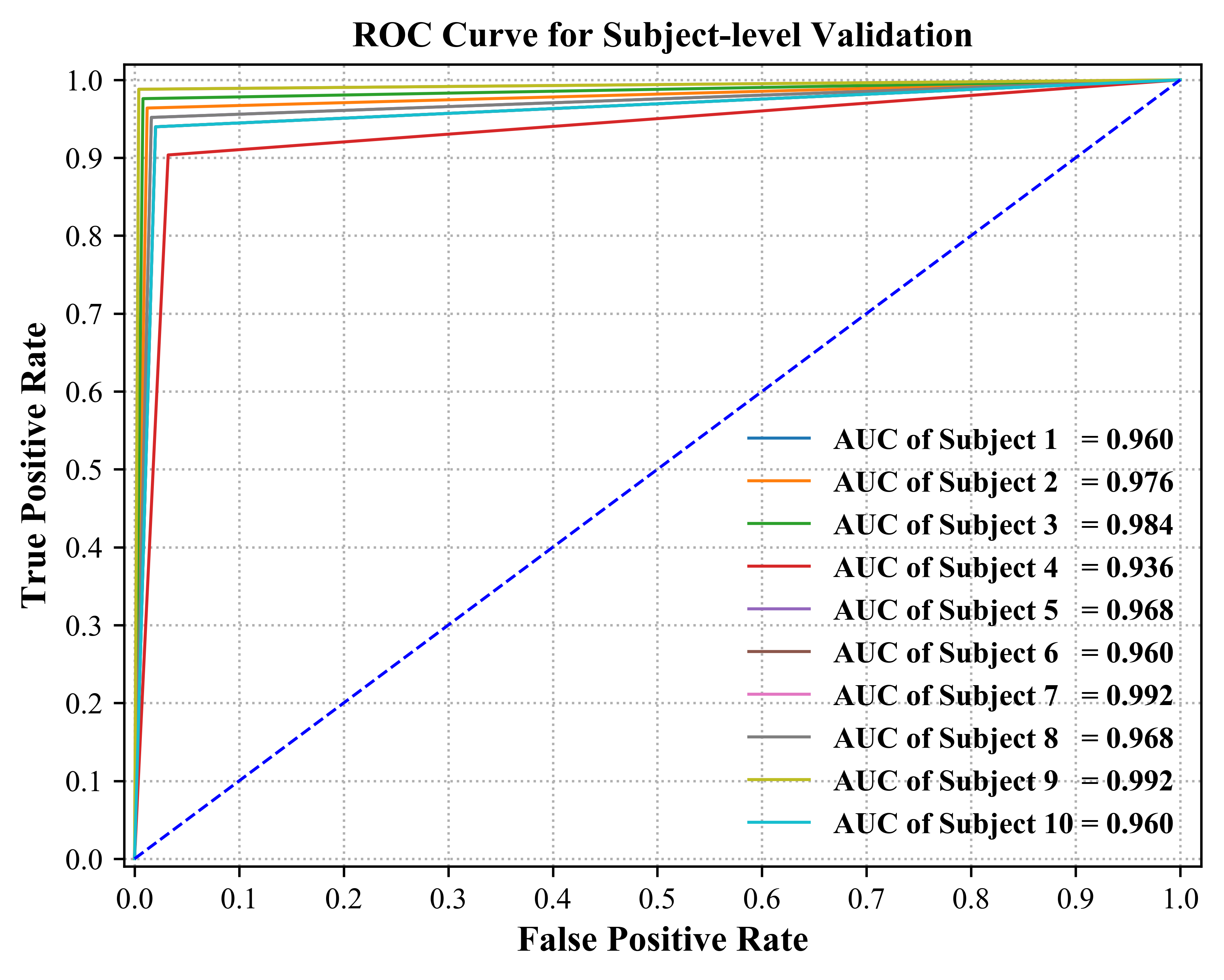}
			\subcaption{ROC Curve w.r.t. Subject-level Validation}
			\label{ROC_Subject_level_Validation}
		\end{minipage}
		\caption{Loss and ROC Curve for subject-level evaluation.}
		\label{Loss_ROC_Subject_Level}
	\end{figure}
	
	As can be seen from \reffig{Loss_Subject_level_Validation}, the model has been shown to converge for the subject-specific adaptation. Receiver Operating Characteristic Curve (ROC Curve) with its corresponding AUC was visible in \reffig{ROC_Subject_level_Validation}.
	
	\begin{table*}[h]
		\centering
		\footnotesize
		\caption{Current studies comparison on subject-level prediction}
		\label{Subject-level-evaluation}
			\begin{tabular}{cccc}
				\hline
				\textbf{Related Work} & \textbf{Max. GAA} & \textbf{Approach} & \textbf{Database} \\ \hline
				Ortiz-Echeverri \emph{et al.} (2019) & 94.66\% & Sorted-fast ICA-CWT + CNNs & \multirow{3}{*}{BCI Competition IV-a Dataset} \\ 
				Sadiq \emph{et al.} (2019) & 95.20\% & EWT + LS-SVM &  \\ 
				Taran \emph{et al.} (2018) & 96.89\% & TQWT + LS-SVM &  \\ 
				Zhang \emph{et al.} (2019) & 83.00\% & CNNs-LSTM & \multirow{3}{*}{BCI Competition IV-2a Dataset} \\ 
				Ji \emph{et al.} (2019) & 95.10\% & SVM &  \\ 
				Amin \emph{et al.} (2019) & 95.40\% & MCNNs &  \\ 
				Dose \emph{et al.} (2018) & 68.51\% & CNNs & \multirow{3}{*}{Physionet Database} \\ 
				Hou \emph{et al.} (2019) & 96.00\% & ESI + CNNs &  \\ 
				\textbf{This work} & \textbf{98.81\%} & \multicolumn{1}{l}{\textbf{Attention-based BiLSTM-GCN}} &  \\ \hline
			\end{tabular}
	\end{table*}

	The comparison of subject-level prediction was put forward between the presented approach and competitive models\cite{ortiz2019new, sadiq2019motor, taran2019motor, zhang2019novel, ji2019eeg, amin2019multilevel, dose2018end, hou2019novel}. The Attention-based BiLSTM-GCN approach has achieved highly accurate results, which suggested robustness and effectiveness towards EEG signal processing as shown in \reftab{Subject-level-evaluation}.
	
	The presented approach has improved classification accuracy, and obtained state-of-the-art results. The reason for the outstanding performance was that the Attention-based BiLSTM model managed to extract relevant features from raw EEG signals. The followed GCN model successfully classified features by cooperating with the topological relationship of overall features. 

\section{Conclusion}\label{Conclusion}
To address the challenge of inter-trial and inter-subject variability in EEG signals, an innovative approach of Attention-based BiLSTM-GCN was proposed to accurately classify four-class EEG MI tasks, i.e., imagining left fist, right fist, both fists, and both feet. First of all, the BiLSTM with Attention model succeeded in extracting relevant features from raw EEG signals. The followed GCN model intensified the decoding performance by cooperating with the internal topological relationship of relevant features, which were estimated from Pearson’s matrix of the overall features. Besides, results provided compelling evidence that the method has converged to both subject-level and group-wise predictions, and achieved the best state-of-the-art performance, i.e., 98.81\% and 94.64\% accuracy, respectively, for handling individual variability, which were far ahead of existing studies. The 0.4-second sample size was proven effective and efficient in prediction compared with traditional 4s trial length, which means that our proposed framework can provide time-resolved solution towards fast response. Results on a group of 20 subjects were derived by 10-fold cross validation indicating repeatability and stability. The proposed method is predicted to advance the clinical translation of the EEG MI-based BCI technology to meet the diverse demands, such as of paralyzed patients. In summary, the unprecedented performance with the highest accuracy and time-resolved prediction were fulfilled via the introduced feature mining approach.

\section*{Acknowledgments}
The authors would like to thank the Brain Team at Google for developing TensorFlow. We further acknowledge PhysioNet for open source the EEG Motor Movement/Imagery Dataset to promote the research. 

\bibliographystyle{ieeetr}
\bibliography{bibliography}

\begin{thebibliography}{10}

\bibitem{bouton2016restoring}
C.~E. Bouton, A.~Shaikhouni, N.~V. Annetta, M.~A. Bockbrader, D.~A.
  Friedenberg, D.~M. Nielson, G.~Sharma, P.~B. Sederberg, B.~C. Glenn, W.~J.
  Mysiw, {\em et~al.}, ``Restoring cortical control of functional movement in a
  human with quadriplegia,'' {\em Nature}, vol.~533, no.~7602, p.~247, 2016.

\bibitem{schwemmer2018meeting}
M.~A. Schwemmer, N.~D. Skomrock, P.~B. Sederberg, J.~E. Ting, G.~Sharma, M.~A.
  Bockbrader, and D.~A. Friedenberg, ``Meeting brain--computer interface user
  performance expectations using a deep neural network decoding framework,''
  {\em Nature medicine}, vol.~24, no.~11, p.~1669, 2018.

\bibitem{daly2008brain}
J.~J. Daly and J.~R. Wolpaw, ``Brain--computer interfaces in neurological
  rehabilitation,'' {\em The Lancet Neurology}, vol.~7, no.~11, pp.~1032--1043,
  2008.

\bibitem{pereira2018eeg}
J.~Pereira, A.~I. Sburlea, and G.~R. M{\"u}ller-Putz, ``Eeg patterns of
  self-paced movement imaginations towards externally-cued and
  internally-selected targets,'' {\em Scientific reports}, vol.~8, no.~1,
  p.~13394, 2018.

\bibitem{mahmood2019fully}
M.~Mahmood, D.~Mzurikwao, Y.-S. Kim, Y.~Lee, S.~Mishra, R.~Herbert, A.~Duarte,
  C.~S. Ang, and W.-H. Yeo, ``Fully portable and wireless universal
  brain--machine interfaces enabled by flexible scalp electronics and deep
  learning algorithm,'' {\em Nature Machine Intelligence}, vol.~1, no.~9,
  pp.~412--422, 2019.

\bibitem{lecun2015deep}
Y.~LeCun, Y.~Bengio, and G.~Hinton, ``Deep learning,'' {\em nature}, vol.~521,
  no.~7553, pp.~436--444, 2015.

\bibitem{lotte2018review}
F.~Lotte, L.~Bougrain, A.~Cichocki, M.~Clerc, M.~Congedo, A.~Rakotomamonjy, and
  F.~Yger, ``A review of classification algorithms for eeg-based
  brain--computer interfaces: a 10 year update,'' {\em Journal of neural
  engineering}, vol.~15, no.~3, p.~031005, 2018.

\bibitem{rumelhart1986learning}
D.~E. Rumelhart, G.~E. Hinton, and R.~J. Williams, ``Learning representations
  by back-propagating errors,'' {\em nature}, vol.~323, no.~6088, pp.~533--536,
  1986.

\bibitem{zhang2018spatial}
T.~Zhang, W.~Zheng, Z.~Cui, Y.~Zong, and Y.~Li, ``Spatial--temporal recurrent
  neural network for emotion recognition,'' {\em IEEE transactions on
  cybernetics}, vol.~49, no.~3, pp.~839--847, 2018.

\bibitem{wang2018lstm}
P.~Wang, A.~Jiang, X.~Liu, J.~Shang, and L.~Zhang, ``Lstm-based eeg
  classification in motor imagery tasks,'' {\em IEEE Transactions on Neural
  Systems and Rehabilitation Engineering}, vol.~26, no.~11, pp.~2086--2095,
  2018.

\bibitem{luo2018exploring}
T.-j. Luo, F.~Chao, {\em et~al.}, ``Exploring spatial-frequency-sequential
  relationships for motor imagery classification with recurrent neural
  network,'' {\em BMC bioinformatics}, vol.~19, no.~1, p.~344, 2018.

\bibitem{guler2005recurrent}
N.~F. G{\"u}ler, E.~D. {\"U}beyli, and I.~G{\"u}ler, ``Recurrent neural
  networks employing lyapunov exponents for eeg signals classification,'' {\em
  Expert systems with applications}, vol.~29, no.~3, pp.~506--514, 2005.

\bibitem{zhang2018mindid}
X.~Zhang, L.~Yao, S.~S. Kanhere, Y.~Liu, T.~Gu, and K.~Chen, ``Mindid: Person
  identification from brain waves through attention-based recurrent neural
  network,'' {\em Proceedings of the ACM on Interactive, Mobile, Wearable and
  Ubiquitous Technologies}, vol.~2, no.~3, p.~149, 2018.

\bibitem{hochreiter1997long}
S.~Hochreiter and J.~Schmidhuber, ``Long short-term memory,'' {\em Neural
  computation}, vol.~9, no.~8, pp.~1735--1780, 1997.

\bibitem{fukushima1980neocognitron}
K.~Fukushima, ``Neocognitron: A self-organizing neural network model for a
  mechanism of pattern recognition unaffected by shift in position,'' {\em
  Biological cybernetics}, vol.~36, no.~4, pp.~193--202, 1980.

\bibitem{lecun1998gradient}
Y.~LeCun, L.~Bottou, Y.~Bengio, P.~Haffner, {\em et~al.}, ``Gradient-based
  learning applied to document recognition,'' {\em Proceedings of the IEEE},
  vol.~86, no.~11, pp.~2278--2324, 1998.

\bibitem{hou2019novel}
Y.~Hou, L.~Zhou, S.~Jia, and X.~Lun, ``A novel approach of decoding {EEG}
  four-class motor imagery tasks via scout {ESI} and {CNN},'' {\em Journal of
  Neural Engineering}, vol.~17, p.~016048, feb 2020.

\bibitem{dose2018end}
H.~Dose, J.~S. M{\o}ller, H.~K. Iversen, and S.~Puthusserypady, ``An end-to-end
  deep learning approach to mi-eeg signal classification for bcis,'' {\em
  Expert Systems with Applications}, vol.~114, pp.~532--542, 2018.

\bibitem{henaff2015deep}
M.~Henaff, J.~Bruna, and Y.~LeCun, ``Deep convolutional networks on
  graph-structured data,'' {\em arXiv preprint arXiv:1506.05163}, 2015.

\bibitem{bruna2014spectral}
J.~Bruna, W.~Zaremba, A.~Szlam, and Y.~Lecun, ``Spectral networks and locally
  connected networks on graphs,'' in {\em International Conference on Learning
  Representations (ICLR2014), CBLS, April 2014}, pp.~http--openreview, 2014.

\bibitem{duvenaud2015convolutional}
D.~K. Duvenaud, D.~Maclaurin, J.~Iparraguirre, R.~Bombarell, T.~Hirzel,
  A.~Aspuru-Guzik, and R.~P. Adams, ``Convolutional networks on graphs for
  learning molecular fingerprints,'' in {\em Advances in neural information
  processing systems}, pp.~2224--2232, 2015.

\bibitem{niepert2016learning}
M.~Niepert, M.~Ahmed, and K.~Kutzkov, ``Learning convolutional neural networks
  for graphs,'' in {\em International conference on machine learning},
  pp.~2014--2023, 2016.

\bibitem{defferrard2016convolutional}
M.~Defferrard, X.~Bresson, and P.~Vandergheynst, ``Convolutional neural
  networks on graphs with fast localized spectral filtering,'' in {\em Advances
  in neural information processing systems}, pp.~3844--3852, 2016.

\bibitem{wang2018eeg}
X.-h. Wang, T.~Zhang, X.-m. Xu, L.~Chen, X.-f. Xing, and C.~P. Chen, ``Eeg
  emotion recognition using dynamical graph convolutional neural networks and
  broad learning system,'' in {\em 2018 IEEE International Conference on
  Bioinformatics and Biomedicine (BIBM)}, pp.~1240--1244, IEEE, 2018.

\bibitem{zhang2019gcb}
T.~Zhang, X.~Wang, X.~Xu, and C.~P. Chen, ``Gcb-net: Graph convolutional broad
  network and its application in emotion recognition,'' {\em IEEE Transactions
  on Affective Computing}, 2019.

\bibitem{song2018eeg}
T.~Song, W.~Zheng, P.~Song, and Z.~Cui, ``Eeg emotion recognition using
  dynamical graph convolutional neural networks,'' {\em IEEE Transactions on
  Affective Computing}, 2018.

\bibitem{wang2019phase}
Z.~Wang, Y.~Tong, and X.~Heng, ``Phase-locking value based graph convolutional
  neural networks for emotion recognition,'' {\em IEEE Access}, vol.~7,
  pp.~93711--93722, 2019.

\bibitem{choetal2014learning}
K.~Cho, B.~van Merri{\"e}nboer, C.~Gulcehre, D.~Bahdanau, F.~Bougares,
  H.~Schwenk, and Y.~Bengio, ``Learning phrase representations using {RNN}
  encoder{--}decoder for statistical machine translation,'' in {\em Proceedings
  of the 2014 Conference on Empirical Methods in Natural Language Processing
  ({EMNLP})}, (Doha, Qatar), pp.~1724--1734, Association for Computational
  Linguistics, Oct. 2014.

\bibitem{bahdanau2014neural}
D.~Bahdanau, K.~Cho, and Y.~Bengio, ``Neural machine translation by jointly
  learning to align and translate,'' {\em arXiv preprint arXiv:1409.0473},
  2014.

\bibitem{xu2015show}
K.~Xu, J.~Ba, R.~Kiros, K.~Cho, A.~Courville, R.~Salakhudinov, R.~Zemel, and
  Y.~Bengio, ``Show, attend and tell: Neural image caption generation with
  visual attention,'' in {\em International conference on machine learning},
  pp.~2048--2057, 2015.

\bibitem{yang2016hierarchical}
Z.~Yang, D.~Yang, C.~Dyer, X.~He, A.~Smola, and E.~Hovy, ``Hierarchical
  attention networks for document classification,'' in {\em Proceedings of the
  2016 conference of the North American chapter of the association for
  computational linguistics: human language technologies}, pp.~1480--1489,
  2016.

\bibitem{chorowski2015attention}
J.~K. Chorowski, D.~Bahdanau, D.~Serdyuk, K.~Cho, and Y.~Bengio,
  ``Attention-based models for speech recognition,'' in {\em Advances in neural
  information processing systems}, pp.~577--585, 2015.

\bibitem{hammond2011wavelets}
D.~K. Hammond, P.~Vandergheynst, and R.~Gribonval, ``Wavelets on graphs via
  spectral graph theory,'' {\em Applied and Computational Harmonic Analysis},
  vol.~30, no.~2, pp.~129--150, 2011.

\bibitem{dhillon2007weighted}
I.~S. Dhillon, Y.~Guan, and B.~Kulis, ``Weighted graph cuts without
  eigenvectors a multilevel approach,'' {\em IEEE transactions on pattern
  analysis and machine intelligence}, vol.~29, no.~11, pp.~1944--1957, 2007.

\bibitem{shi2000normalized}
J.~Shi and J.~Malik, ``Normalized cuts and image segmentation,'' {\em
  Departmental Papers (CIS)}, vol.~22, no.~8, pp.~888--905, 2000.

\bibitem{srivastava2014dropout}
N.~Srivastava, G.~Hinton, A.~Krizhevsky, I.~Sutskever, and R.~Salakhutdinov,
  ``Dropout: a simple way to prevent neural networks from overfitting,'' {\em
  The journal of machine learning research}, vol.~15, no.~1, pp.~1929--1958,
  2014.

\bibitem{ioffe2015batch}
S.~Ioffe and C.~Szegedy, ``Batch normalization: Accelerating deep network
  training by reducing internal covariate shift,'' {\em arXiv preprint
  arXiv:1502.03167}, 2015.

\bibitem{hahnloser2000digital}
R.~H. Hahnloser, R.~Sarpeshkar, M.~A. Mahowald, R.~J. Douglas, and H.~S. Seung,
  ``Digital selection and analogue amplification coexist in a cortex-inspired
  silicon circuit,'' {\em Nature}, vol.~405, no.~6789, p.~947, 2000.

\bibitem{kingma2014adam}
D.~P. Kingma and J.~Ba, ``Adam: A method for stochastic optimization,'' {\em
  arXiv preprint arXiv:1412.6980}, 2014.

\bibitem{zhang2004solving}
T.~Zhang, ``Solving large scale linear prediction problems using stochastic
  gradient descent algorithms,'' in {\em Proceedings of the twenty-first
  international conference on Machine learning}, p.~116, ACM, 2004.

\bibitem{abadi2016tensorflow}
M.~Abadi, P.~Barham, J.~Chen, Z.~Chen, A.~Davis, J.~Dean, M.~Devin,
  S.~Ghemawat, G.~Irving, M.~Isard, {\em et~al.}, ``Tensorflow: A system for
  large-scale machine learning,'' in {\em 12th $\{$USENIX$\}$ Symposium on
  Operating Systems Design and Implementation ($\{$OSDI$\}$ 16)}, pp.~265--283,
  2016.

\bibitem{goldberger2000physiobank}
A.~L. Goldberger, L.~A. Amaral, L.~Glass, J.~M. Hausdorff, P.~C. Ivanov, R.~G.
  Mark, J.~E. Mietus, G.~B. Moody, C.-K. Peng, and H.~E. Stanley, ``Physiobank,
  physiotoolkit, and physionet: components of a new research resource for
  complex physiologic signals,'' {\em Circulation}, vol.~101, no.~23,
  pp.~e215--e220, 2000.

\bibitem{tanaka2020group}
H.~Tanaka, ``Group task-related component analysis (gtrca): a multivariate
  method for inter-trial reproducibility and inter-subject similarity
  maximization for eeg data analysis,'' {\em Scientific Reports}, vol.~10,
  no.~1, pp.~1--17, 2020.

\bibitem{ma2018improving}
X.~Ma, S.~Qiu, C.~Du, J.~Xing, and H.~He, ``Improving eeg-based motor imagery
  classification via spatial and temporal recurrent neural networks,'' in {\em
  2018 40th Annual International Conference of the IEEE Engineering in Medicine
  and Biology Society (EMBC)}, pp.~1903--1906, IEEE, 2018.

\bibitem{ortiz2019new}
C.~J. Ortiz-Echeverri, S.~Salazar-Colores, J.~Rodr{\'\i}guez-Res{\'e}ndiz, and
  R.~A. G{\'o}mez-Loenzo, ``A new approach for motor imagery classification
  based on sorted blind source separation, continuous wavelet transform, and
  convolutional neural network,'' {\em Sensors}, vol.~19, no.~20, p.~4541,
  2019.

\bibitem{sadiq2019motor}
M.~T. Sadiq, X.~Yu, Z.~Yuan, Z.~Fan, A.~U. Rehman, G.~Li, and G.~Xiao, ``Motor
  imagery eeg signals classification based on mode amplitude and frequency
  components using empirical wavelet transform,'' {\em IEEE Access}, vol.~7,
  pp.~127678--127692, 2019.

\bibitem{taran2019motor}
S.~Taran and V.~Bajaj, ``Motor imagery tasks-based eeg signals classification
  using tunable-q wavelet transform,'' {\em Neural Computing and Applications},
  vol.~31, no.~11, pp.~6925--6932, 2019.

\bibitem{zhang2019novel}
R.~Zhang, Q.~Zong, L.~Dou, and X.~Zhao, ``A novel hybrid deep learning scheme
  for four-class motor imagery classification,'' {\em Journal of neural
  engineering}, vol.~16, no.~6, p.~066004, 2019.

\bibitem{ji2019eeg}
N.~Ji, L.~Ma, H.~Dong, and X.~Zhang, ``Eeg signals feature extraction based on
  dwt and emd combined with approximate entropy,'' {\em Brain sciences},
  vol.~9, no.~8, p.~201, 2019.

\bibitem{amin2019multilevel}
S.~U. Amin, M.~Alsulaiman, G.~Muhammad, M.~A. Bencherif, and M.~S. Hossain,
  ``Multilevel weighted feature fusion using convolutional neural networks for
  eeg motor imagery classification,'' {\em IEEE Access}, vol.~7,
  pp.~18940--18950, 2019.

\end{thebibliography}
\end{document}